\documentclass[a4paper,fleqn,usenatbib]{mnras}

\usepackage[T1]{fontenc}
\usepackage{ae,aecompl}
\usepackage[switch,modulo]{lineno}

\usepackage{graphicx}	
\usepackage{amsmath}	
\usepackage{amssymb}	

\usepackage[caption = false]{subfig} 

\title[MUSE view of NGC 5626]{MUSE stares into the shadows: the high-resolution dust attenuation curve of NGC 5626}
\author[S. Viaene et al.]{
S. Viaene$^{1,2}$\thanks{E-mail: sebastien.viaene@ugent.be}, 
M. Sarzi$^{2}$, 
M. Baes$^{1}$,
J. Fritz$^{3}$,
I. Puerari$^{4}$
\\
$^{1}$Sterrenkundig Observatorium, Universiteit Gent, Krijgslaan 281, B-9000 Gent, Belgium\\
$^{2}$Centre for Astrophysics Research, University of Hertfordshire, College Lane, Hatfield AL10 9AB, UK \\
$^{3}$Instituto de Radioastronom\'{i}a y Astrof\'{i}sica (IRyA-UNAM), Antigua Carrettera a
P\'{a}tzcuaro, 8701, Morelia, Michoac\'{a}n, Mexico. \\
$^{4}$Instituto Nacional de Astrof\'{i}sica, Optica y Electr\'{o}nica, Calle Luis Enrique Erro 1, 72840 Santa Mar\'{i}a Tonantzintla, Puebla, Mexico
}

\date{Accepted XXX. Received YYY; in original form ZZZ}

\pubyear{2017}

\begin{document}
\label{firstpage}
\pagerange{\pageref{firstpage}--\pageref{lastpage}}
\maketitle

\begin{abstract}
The newest generation of integral field unit spectrographs brings three-dimensional mapping of nearby galaxies one step closer. While the focus up to this point was mostly on stars and ionized gas, it is also possible to look at dust in a new, more complete way. Using MUSE science verification observations of NGC 5626, we map the interstellar matter in this dusty lenticular. We use the resolving power of MUSE to measure the optical attenuation with a spectral resolution of 6.25 \AA, at physical scales of $0.1-1$ kpc. 
The integrated attenuation curve of NGC 5626 shows a smooth, slightly steeper than Milky Way and SMC attenuation curves. Several sharp features are superimposed: we measure lower attenuation at spectral emission lines and higher attenuation for the sodium line doublet. No correlation was observed between sodium line strength and reddening by dust on spatially resolved scales. Additionally, the continuum attenuation was found to be independent from the Balmer decrement (tracing ionized gas attenuation).
We model and interpret the variations in the attenuation curves of each spatial resolution element of NGC 5626. We find that the amount and distribution of dust along the line-of-sight is highly degenerate with any variation in the intrinsic extinction law. Our analysis shows that the interstellar matter in NGC 5626 resides in a regular and well-settled disk. Our results preach caution in the application of simple recipes to de-redden global galaxy spectra and underlines the need for more realistic dust geometries when constructing such correction formulas.
\end{abstract}




\section{Introduction}

Dust in galaxies is a vital ingredient for an efficient transformation of gas into stars. It facilitates molecular cloud formation and regulates the gas temperature in the interstellar medium (ISM). While the broad effect of dust in galaxies is well-known, many uncertainties lie in the chemical composition, grain size distribution, and optical properties \citep[see e.g.][]{Draine2003, Jones2017}. This translates in uncertainties when correcting observations for the effects of dust. Indeed, one of the main effects of dust for the observer is the obscuration of ultraviolet (UV), optical and near-infrared (NIR) light. On average, dust converts about one third of the emitted stellar energy into far-infrared and sub-millimeter emission \citep[e.g.][]{Popescu2002, Skibba2011, Viaene2016}. A proper quantification for the effects of dust is necessary for key galaxy measurements such as luminosities and colours, surface brightness profiles, the initial mass function, and star formation rates \citep[see e.g. ][]{Byun1994,Mollenhoff2006,Gadotti2010}.

Quantifications of dust effects are often formulated as extinction laws \citep[see e.g.][]{Mathis1990}. An extinction law is an intrinsic property of a given dust mix. It reflects, as a function of wavelength, the fraction of the radiation that is absorbed or scattered, per unit dust mass. In our own Milky Way (and in the Local Group), extinction laws can be measured by comparing the flux between unobscured and obscured stars of the same spectral type \citep[see e.g.][]{Gordon2003,Fitzpatrick2007, Clayton2015}.
In galaxies further away, individual stars cannot be resolved. Along the line of sight, a mix of stars and dust creates a complex, \textit{effective} extinction law, often called an attenuation curve. We refer the reader to \citet{Krugel2009} for a comprehensive, theoretical, discourse on intrinsic and effective extinction. Practically, attenuation curves contain not only absorption and scattering out of the line of sight (as goes for extinction laws), but also scattering of light into the line of sight, and direct (unobscured) light. This makes the attenuation curve a convolution of the intrinsic dust extinction law and the star-dust distribution along the line of sight. 

In principle, one needs to use a suitable attenuation law to correct UV/optical/NIR observations for the effects of dust. Only then the intrinsic  stellar emission is recovered and can it be examined. In practice, however, several general recipes are available, such as average galaxy attenuation laws \citep[e.g.][]{Calzetti1994, Wild2011, Battisti2016}, attenuation laws from resolved studies \citep[e.g.][]{Liu2013, Kreckel2013, Dong2014}, or even pure extinction laws \citep[e.g. ][]{Gordon2003, Fitzpatrick2007, Clayton2015}.

There are three big caveats to these approaches:
First, locally derived extinction laws (MW, SMC, LMC) are usually an average of different lines of sight, potentially blending different dust mixtures.
Second, when starting from a locally derived extinction law, one assumes that the dust in the studied galaxy has the same properties as its local counterpart (MW, SMC, LMC).
Third, extinction laws and attenuation curves derived from them usually assume that the dust is distributed in a thin screen between the stars and the observer. This is a huge simplification. 
Finally, applying one correction curve assumes that the dust has the same properties across the entire galaxy. 
In this paper we examine whether we can measure attenuation curves in a galaxy beyond the local group, and how this should be interpreted in terms of spatial variations of the dust properties.

Dusty early-type galaxies (ETGs) are excellent laboratories to study dust on a galaxy-wide scale because their stellar body is  regular and fairly easy to model \citep{Ferrarese2006, DeLooze2012a, Viaene2015, Cappellari2016}. They are also an important subclass to understand the evolution from late-type galaxies (LTGs) to ETGs.
In the classic evolutionary scenario, an LTG transitions from `active' to `passive' due to rapid quenching of their star formation \citep[e.g.][and references therein]{Ciesla2016}. Recently, however, \citet{Eales2017}, argued that the strict segregation between active and passive galaxies is a mere observational effect. They found a curved but continuous galaxy evolutionary sequence using the Herschel Reference Survey \citep{Boselli2010}. In the specific SFR vs. stellar mass plot, the galaxies that make the smooth connection between active and passive galaxies are mostly ETGs holding interstellar matter (gas and dust) and exhibiting signs of star formation. Characterizing the morphology of dust and ionized gas in these systems could therefore bring clues to the way these dusty ETGs formed and what their role is in the general galaxy evolution sequence.

\begin{figure*}
	\includegraphics[width=\textwidth]{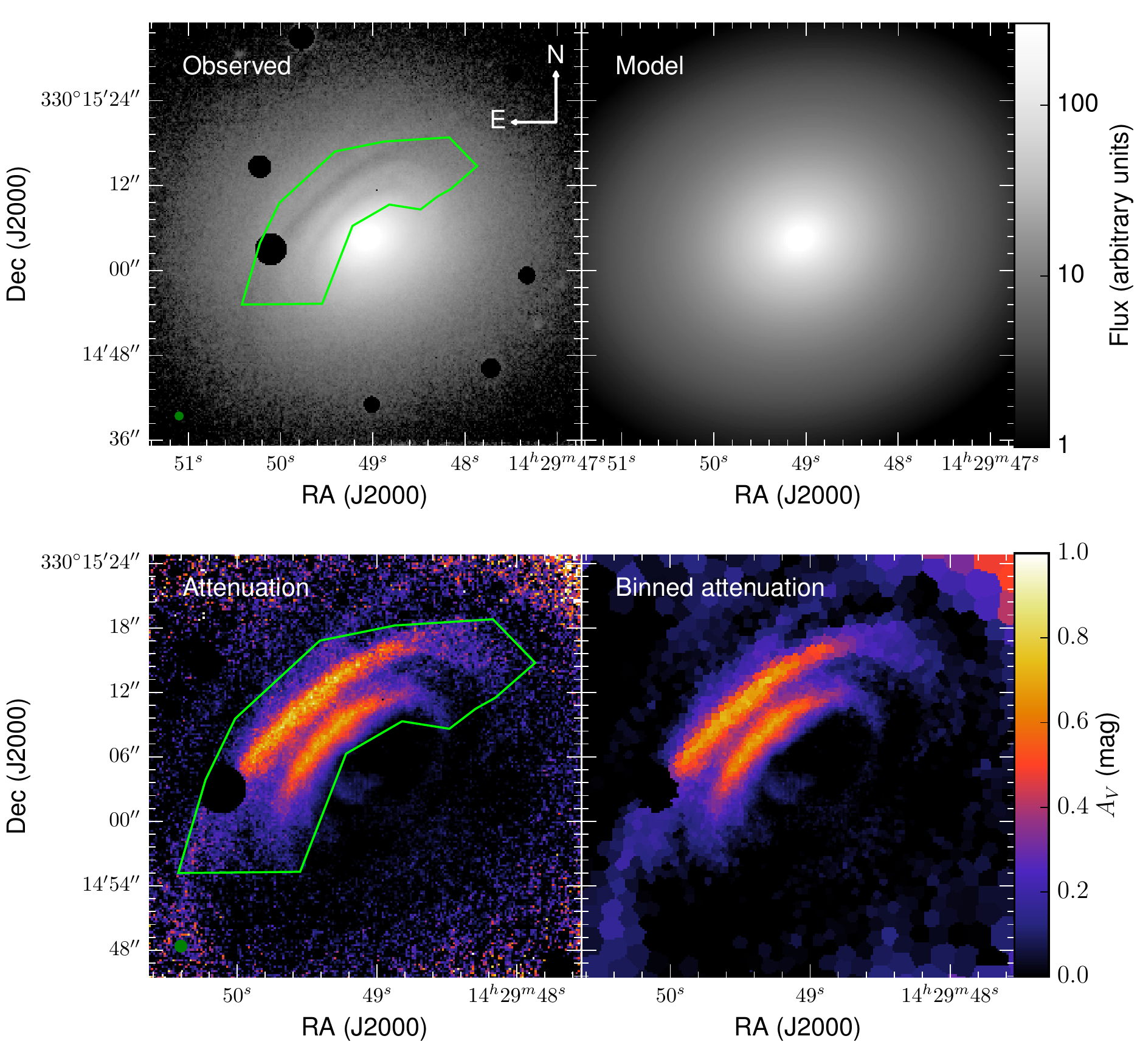}
    \caption{Different stages of the MGE modelling to achieve maps of the attenuation (see Sect.~\ref{sec:MGEmodeling}). Top left: observed slice of the sky-subtracted datacube. The slice has a rest-frame wavelength of 5503.75 \AA\ and a width of 6.25 \AA. The brighter foreground stars are masked out. The area of influence of the dust lane area is indicated by the green region. The green circle in the lower left corner represents the size of the PSF. Top right: Dust free model constructed by fitting the dust-free regions of the observed slice. This MGE model uses 5 Gaussians to constrain the light profile. Bottom left: attenuation map, obtained by subtracting the dust-free model from the observed image. Bottom right: Voronoi-binned attenuation map. The maps is the bottom line are zoomed in.}
    \label{fig:MGEexample}
\end{figure*}

In this paper, we explore the dust and ionized gas in NGC 5626, a lenticular galaxy at redshift $0.023$ with remarkable dust lanes (\citealt{Hawarden1981, Finkelman2008}; Fig.~\ref{fig:MGEexample} top left panel). The subclass of dust-lane ETGs was invoked by \citet{Bertola1978} and sparked studies of its attenuation curves \citep{Brosch1985, Brosch1990, Goudfrooij1994b, Sahu1998, Patil2007, Finkelman2008, Finkelman2010, Kulkarni2014}.

These systems proved quite suitable to determine attenuation curves because the dusty regions can be easily compared to the dust-free areas. The underlying assumption is that the stellar surface brightness profile is regular and predictable across the galaxy. Previous measurements are based on broad band ($UBVRI$) observations, producing a handful of data points in the attenuation vs. wavelength plane. It provides sufficient accuracy to compute the so-called $R_V$ value, which is the ratio of total attenuation in the V band, to the selective $A_B-A_V$ attenuation. $R_V$ values typically range from $2$ to $4$ \citep{Patil2007, Finkelman2008}, and are - on average - in line with $R_V = 3.1$ for the Milky Way \citep{Valencic2004, Fitzpatrick2007}. This suggests that the dust in dusty ETGs is similar to that of the Milky Way.

Recently, with the advent of integral field unit (IFU) spectrographs, we can resolve galaxies both spatially and spectrally. This has revolutionised our understanding of ETGs and their ISM content \citep[see][for a review]{Cappellari2016}. 

Today, the second generation of IFU spectrographs are more sensitive and provide higher spatial and spectral resolution (e.g. MUSE; \citealt{MUSE}) or mass applicability (e.g. SAMI; \citealt{SAMI} and MaNGA; \citealt{MaNGA}). This paves the road towards 3D dissection of all galaxy components. We will perform a detailed study of NGC 5626 using MUSE observations. This allows us to study dust attenuation in unprecedented spectral resolution for galaxies, and simultaneously map the ionized gas in this system. While the focus of this paper lies on the ISM, a follow-up study will target the stellar kinematics of this object, and model the dust-starlight energy balance using radiative transfer simulations.

In Sect.~\ref{sec:MUSE} the observations with MUSE and the data reduction are described. Sect.~\ref{sec:analysis} explains the attenuation curve measurements and interpretation on both integrated and spatially resolved scale. Additionally, we look at spectral features and their connection to reddening by dust. A summary with the main conclusions of this investigation is presented in Sect.~\ref{sec:conclusions}.

\section{Data acquisition} \label{sec:MUSE}

Optical observations of NGC 5626 were obtained with the Multi Unit Spectroscopic Explorer (MUSE; \citealt{MUSE}) instrument. This integral field unit (IFU) is commissioned at unit telescope 4 (UT4) of the Very Large Telescope (VLT). MUSE has a spectral range of 4750-9350 \AA, a spectral sampling of 1.25 \AA\ ($R=1770-3590$), and a pixel scale of 0.2 arcseconds. The wide-field mode yields a $1\times1$ arcmin field of view. These combined capabilities allow for the most high-resolution IFU observations to date.

The observations were obtained in the frame of the science verification campaign during the nights of 22-25th of August, 2014 (run 60.A-9325(A)). The seeing was found to be 1.0 arcsecond. The first exposure failed and had to be re-done. In total, 4 exposures of 1800s on target were made, each time with a relative rotation of 90 degrees. The latter procedure was advised to obtain a dithering effect.

The Science Verification data were processed following the recommendations in the standard pipeline, version 1.0. This includes: bias and flat field correction, wavelength calibration for each IFU, measurement of the line spread function, illumination correction to obtain a basic science image. In a second phase, the basic science frames were flux calibrated and astrometry coordinates were added. Furthermore, a sky spectrum was created to correct for the emission of the sky as much as possible. We tried multiple ways to correct for the sky emission. In the end, the \textit{``subtract\_model"} option for the \textit{skymethod} setting of \textit{muse\_scipost} was able to reduce the contaminating sky lines the most, without over-subtracting flux or introducing background gradients. We will perform a second sky correction prior to our analysis (see below).

From each of the four individual, dithered exposures, a white light image was made. These images were used to compute offsets in WCS to ensure a proper alignment. The four individual exposures were then combined using the standard \textit{muse\_scipost} routine, ensuring an enhanced signal-to-noise ratio.

The final datacube still contains a number of strong emission lines from the night sky, especially in the red part of the spectra. For our purposes, we chose to perform a secondary sky-subtraction, exploiting the 3D nature of the datacube. We treat each wavelength slice as an independent image, with NGC 5626 in the center of the field of view. A simple inspection of the surface brightness profile indicated that the surface brightness of NGC 5626 drops well below the sky noise level in the outskirts of the image. This means we can use the corners of the image to estimate the sky level, without introducing large biases in the surface brightness profile of the galaxy itself. The four corners of the field of view were thus used to estimate the sky level for each wavelength slice. This natural way of sky correction proved to be a simple and consistent way to correct for many of the sky features. 

There are a few remaining lines in the sky-subtracted datacube that could not be corrected for. These will show up as very narrow lines in our attenuation curves. We will return to them later in our analysis.

\section{Analysis} \label{sec:analysis}

NGC 5626 shows strong signatures of dust extinction North-East from its center. The first step in our analysis consists of measuring the attenuation (or effective extinction) in each resolution element of this galaxy. One can essentially see the MUSE datacube as a series of very narrow band images of NGC 5626. We will exploit this by treating each image individually, and create a dust-free model. To do so, we first bin the spectral slices by a factor of  five to boost the SNR. This creates 736 slices of $6.25$ \AA\ in wavelength width. They form the base of our analysis.

\subsection{Dust-free galaxy model} \label{sec:MGEmodeling}

To construct a dust-free model for each slice, we first mask out bright foreground stars. Next, we determine the regions most affected by dust attenuation. This was done by plotting surface brightness contours and mark the area where they bend inwards, clearly deviating from an elliptical shape (green polygon in Fig.~\ref{fig:MGEexample}). What remains are the areas of the galaxy without extinction from the dust lane, and without foreground emission.

We note that these regions can be reddened as well, by a diffuse dust component inside the galaxy, or by foreground Galactic dust. The latter will not influence our results as we work with the relative extinction between observations and a dust-free model of the galaxy. In other words, the measurement of the attenuation law is self-consistent and unaffected by Galactic dust. A reservoir of diffuse dust was long suggested to explain the discrepancy between dust masses derived from optical broad band data, and those from FIR dust emission \citep[e.g.][]{Goudfrooij1995, Patil2007, Finkelman2008, Finkelman2010}. However, kinematical analyses \citep{Vanderbeke2011}, observed colour gradients \citep{Michard2005}, and radiative transfer models \citep{Viaene2015} suggest that a diffuse dust component in ETGs is rather unlikely. In the few cases where FIR dust emission of dust-lane ETGs is spatially resolved \citep[e.g. ][]{DeLooze2012a, Dale2012}, no evidence was found for an extended diffuse dust component either.

In short, we use the dust-free areas of the galaxy to construct a dust-free model for the full galaxy. The difference between this model and the observed slice then yields a map of the attenuation in the dust lane. To model the dust-free regions of each slice, we use Multi-Gaussian Expansion fitting (MGE), using the code by \citet{Cappellari2002}. This method proved to work well in our latest analysis of NGC 4370 \citep{Viaene2015}, a comparable dust ETG. MGE basically adds several 2D Gaussian distributions to mimic the surface brightness profile of a galaxy. We leave the position angle of each Gaussian function as a free parameter. This is necessary to capture deviations from an elliptical surface brightness profile, also referred to as the boxiness of a galaxy \citep[see][for a detailed explanation]{Viaene2015}.

To prevent overfitting, we fix the center of the galaxy at $\alpha$ = 14:29:49.087, $\delta$ = -29:44:55.658. Additionally, the number of Gaussians was varied from 3 to 15, and the effect on the fit was evaluated through the reduced $\chi^2$. We found that five Gaussians are sufficient to capture the surface brightness profile of the galaxy and fix this number for each slice. A smaller number of Gaussians yielded higher $\chi^2$ values, while using more than five Gaussians did not lower the $\chi^2$ significantly. To fit each slice, the free parameters are the magnitude, width, flattening and position angle for each Gaussian. This amounts to 20 free parameters, which are constrained by over 1200 flux measurements in each image.

The various steps of MGE fitting are shown in figure~\ref{fig:MGEexample} for the slice centred at $5503.75$ \AA, corresponding to a classic $V$ band observation. The top panels show the observed slice and the dust-free model. The bottom left panels zoom in on the corresponding attenuation maps. The attenuation map was computed by subtracting observation ($I_{\lambda\mathrm{,obs}}$) and model ($I_{\lambda\mathrm{,model}}$) in log space:
\begin{equation} \label{eq:mgeatt}
A_\lambda = -2.5\log\left(\frac{I_{\lambda\mathrm{,obs}}}{I_{\lambda\mathrm{,model}}}\right),
\end{equation}
The bottom right panel of Fig.~\ref{fig:MGEexample} shows the same attenuation map, but in a Voronoi binning. The choice of the Voronoi bins are based on a full IFU spectral fit with GandALF \citep{Sarzi2006}. We will present this fit along with an in-depth dynamical analysis of this galaxy in our follow-up paper. For the current study, it is sufficient to note that the Voronoi binning is chosen to obtain the best possible signal-to-noise ratio (SNR) for as many bins as possible. Hence in the inner regions, the Voronoi cells are smallest, up to a single pixel. In the outer regions, the bins can contain several hundreds of pixels. In each bin, the attenuation values of the individual pixels are averaged and the  standard deviation reflects the uncertainty of $A_\lambda$ in each bin.

\subsection{Integrated attenuation curve}

\begin{figure*}
	\includegraphics[width=\textwidth]{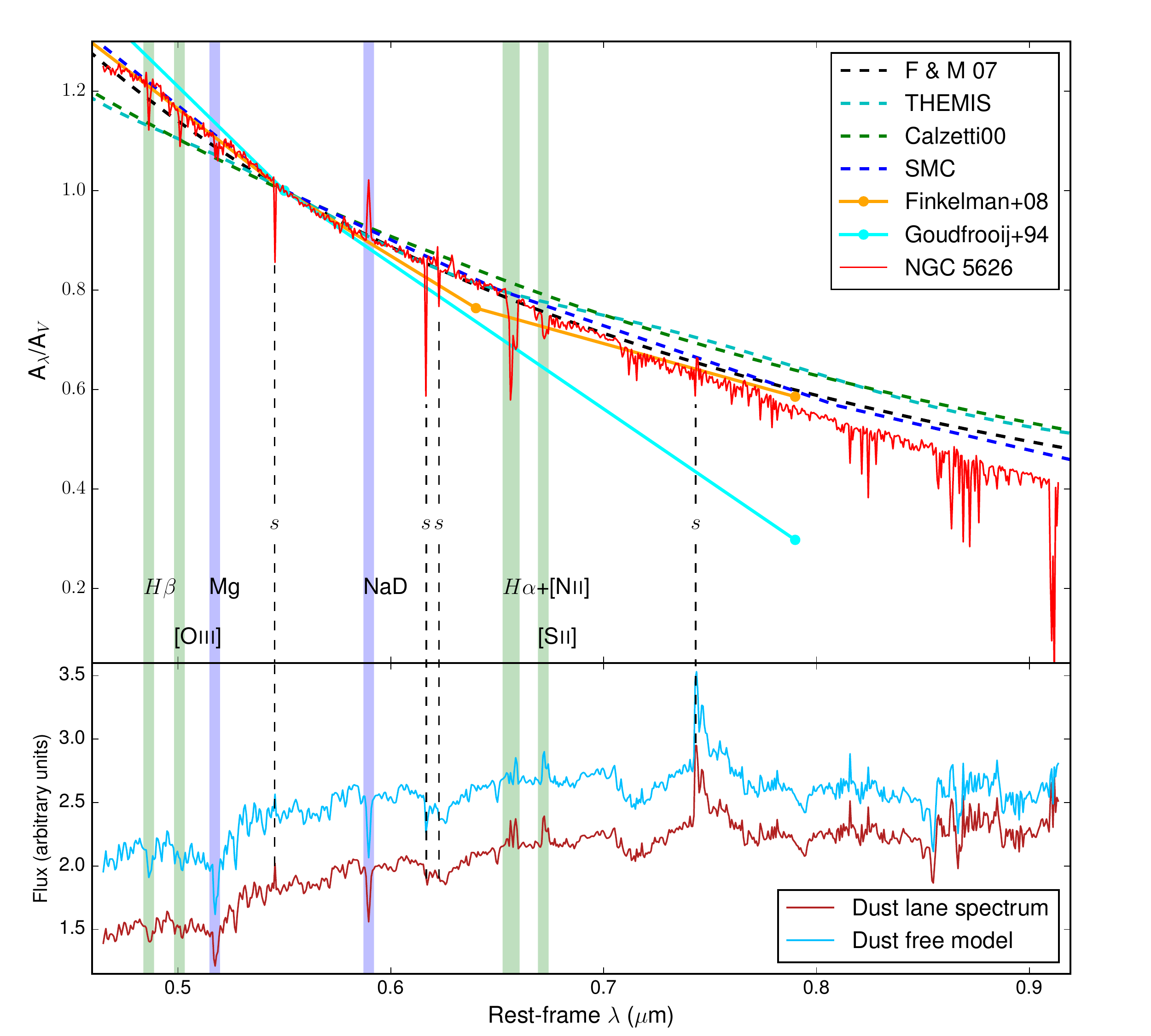}
    \caption{Top panel: high-resolution optical attenuation curve of NGC 5626 (red line), normalised to the $V$ band value. Several observational and theoretical extinction curves are plotted as well for comparison. In particular, attenuation curves for the same galaxy are shown in orange \citep{Finkelman2008} and cyan \citep{Goudfrooij1994b}. The attenuation curve is generally steeper than the Milky Way, and follows the SMC curve most closely. Bottom panel: integrated emission spectrum of the dust-lane region (green polygon in Fig.~\ref{fig:MGEexample}) of NGC 5626 in red. The corresponding dust-free spectrum for this region is shown in blue and based on the MGE fitting of each image slice. Several spectral features are indicated with shaded green (emission) and blue (absorption) areas. Residual sky lines are indicated with vertical dashed lines.}
    \label{fig:att_norm}
\end{figure*}

\subsubsection{Attenuation at spectral resolution} \label{sec:globalAtt}

To obtain the global attenuation curve of NGC 5626, we measure the flux of both observation and model inside the dust lane. This was done for each slice, applying equation~\ref{eq:mgeatt}. The global attenuation curve is plotted in the top panel of Fig.~\ref{fig:att_norm}. To our knowledge, this is the first truly spectral-resolution (6.25 \AA) measurement of a spatially resolved optical attenuation curve of an \textit{individual} galaxy. Previous optical spectroscopic results \citep[e.g.][]{Calzetti1994, Battisti2016, Battisti2017} are based on global, unresolved measurements. They combine large samples of galaxies together to infer attenuation properties whereas MUSE allows for an in-depth individual approach.

\begin{figure*}
	\includegraphics[width=1.0\textwidth]{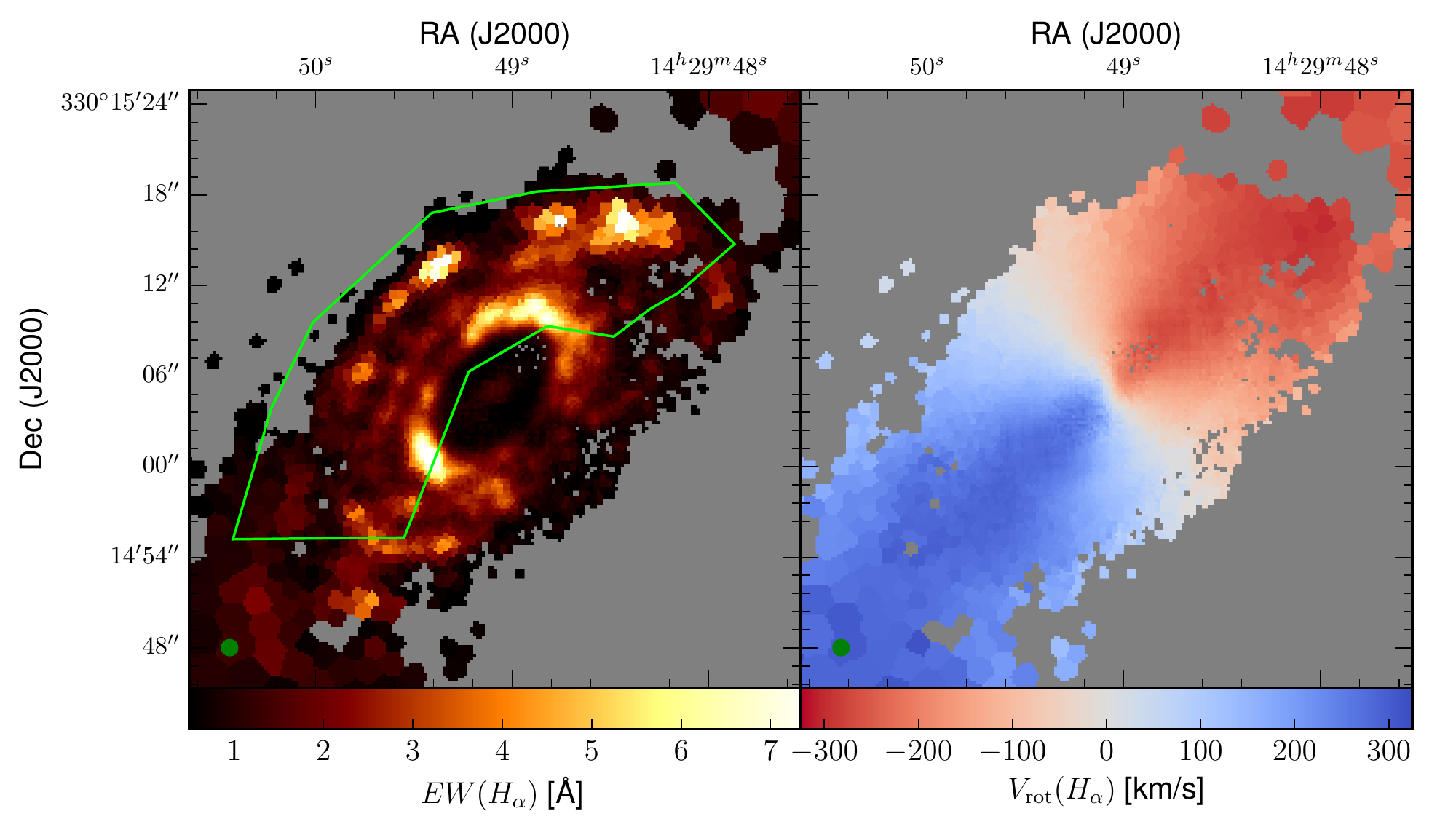}
    \caption{Ionized gas in NGC 5626 as traced by the H$\alpha$ line. Left: Equivalent width $EW (H\alpha)$ revealing multi-ring morphology with hints of spiral arms in the outer regions. The area of influence of the dust lane area is indicated by the green region (see also Fig.~\ref{fig:MGEexample}). Right: Rotational velocity map $V_\mathrm{rot} (H\alpha)$ showing a remarkably regular velocity field. The PSF size is indicated in green in the lower left corner.}
    \label{fig:HaMaps}
\end{figure*}

We compare it with three reference extinction laws: a) The average Milky Way extinction curve from \citet{Fitzpatrick2007}, b) the SMC extinction curve from \citet{Gordon2003}, c) The THEMIS theoretical extinction curve, calibrated for MW dust \citep{Ysard2015, Jones2017}. The core of this extinction law is a physically motivated dust mixture of amorphous (hydro-)carbons and silicates with iron nano-inclusions. It is consistent with the available laboratory data \citep{Jones2013} and explains the variations in diffuse Milky Way dust emission measured by the Planck mission \citep[see][for further details]{Ysard2015}. Additionally, we show the starburst attenuation law described in \citet{Calzetti2000}, which will be used for further modelling (see Sect.~\ref{sec:modelling}). 

It is never straightforward to compare an attenuation curve (which includes extinction, multiple scattering and geometrical effects), with an extinction curve (which only includes absorption and scattering out of the line-of-sight). We address this issue in more depth during our resolved analysis in Sect.~\ref{sec:fitAcurves}. Nevertheless, when normalizing both types of reddening curves to their respecitive $V$ band value, one can qualitatively compare their slope. We compare these curves as a sanity check for our wavelength-slice based attenuation measurement. Our attenuation curve is slightly steeper, but closely resembles the general declining pattern, which is reassuring. Qualitatively, the attenuation curve in NGC 5626 most resembles the SMC attenuation curve.

The attenuation curve in NGC 5626 was measured previously by \citet{Goudfrooij1994b} and \citet{Finkelman2008}. We also show these broad band measurements in Fig.~\ref{fig:att_norm}. Our measurement is consistent with the attenuation curve of \citet{Finkelman2008}, but flatter than \citet{Goudfrooij1994b}. However, the coarse spatial (FWHM $>2$ arcsec) and spectral resolution (3 broad band filters) of their datasets make for a difficult comparison.

There are various features visible in the high-resolution attenuation curve. As an aid to interpret the nature of these features, we plot the mean spectrum of the dust lane pixels in the bottom panel of Fig.~\ref{fig:att_norm}. We chose not to mask out any of the strong lines in the attenuation curve since they will not hinder the determination of the slope or total level of attenuation in a significant way.

First, we note several strong features related to sky emission and absorption. This results in the sharp dip (sometimes followed by a sharp peak) in the attenuation curve at 5454, 6167, 6227 \AA. Furthermore, the broad emission feature in the spectrum at 7433 \AA\ is due to sky-emission that we were unable to filter out. In fact, the MUSE spectra are notoriously populated with sky-lines beyond 7400 \AA. Fortunately, we can still use the continuum emission at these longer wavelengths to reliably measure the attenuation. This is evident from the smooth continuation of the main slope of the attenuation curve as it goes from short to longer wavelengths. We do note that the attenuation curve is noisier at the reddest wavelengths due to the influence of these sky-lines. 

Secondly, there are clear signatures of H$\beta$, [O\textsc{iii}], Mg, H$\alpha$, [N\textsc{ii}] and [S\textsc{ii}] visible as dips in the attenuation curve. Inversely, the sodium doublet (NaD) shows up as a strong peak in attenuation. As a caveat to the signatures associated to the emission lines, we recognize that here, the MGE method is approaching its limit. The line emission (especially H$\alpha$) in NGC 5626 is clumpy and not symmetric. This may challenge our assumption that the light behind the dust line can be extrapolated from the dust-free part of the galaxy. However, this effect can only partly explain the decrease in attenuation as the flux at these line emission wavelengths is still dominated by the continuum emission. Note that the above caveat is only relevant for emission lines. Absorption lines (e.g. NaD) are not affected by this.

Through H$\alpha$, the brightest emission line across the galaxy, we are able to trace the ionized gas. In contrast to the dust (in extinction), we can trace this ISM component in emission across the entire galaxy. The map of equivalent width EW(H$\alpha)$ in Fig.~\ref{fig:HaMaps} reveals a multi-ring morphology. There is not much ionized gas in the center and outskirts. On the North-East side the regions of higher EW(H$\alpha)$ coincide with the arcs of dust extinction. This is expected, as dust and gas usually share a common location in early-type galaxies \citep[see e.g.][]{Sarzi2006, Finkelman2010, Davis2011}. Interestingly, the velocity field of the ionized gas, $V_\mathrm{rot}$ (H$\alpha$), is remarkably regular (see Fig.~\ref{fig:HaMaps}, right panel). This suggests that the ionized gas has settled in a regular disk of low vertical scale height. 
 
Third (in the integrated attenuation curve), we looked for signatures of diffuse interstellar bands (DIBS, see \citealt{Herbig1995} and \citealt{Sarre2006} for reviews). These mysterious non-stellar absorption features occur predominantly in spectra of reddened stars or galaxies. It is still unclear what the carriers are, although they are thought to be of molecular origin, and carbon-bearing. Good candidates are Polycyclic Aromatic Hydrocarbons (PAHs) or fullerenes, although clear evidence for this is still lacking \citep[see e.g.][ and references therein]{Sarre2006, Snow2014, Baron2015, MonrealIbero2015}. Most recently, two DIBs have been attributed to C$_{60}^+$ in laboratory environment \citep{Campbell2015}. Unfortunately, these DIBs occur at longer wavelengths than covered by MUSE.

\begin{figure}
	\includegraphics[width=0.45\textwidth]{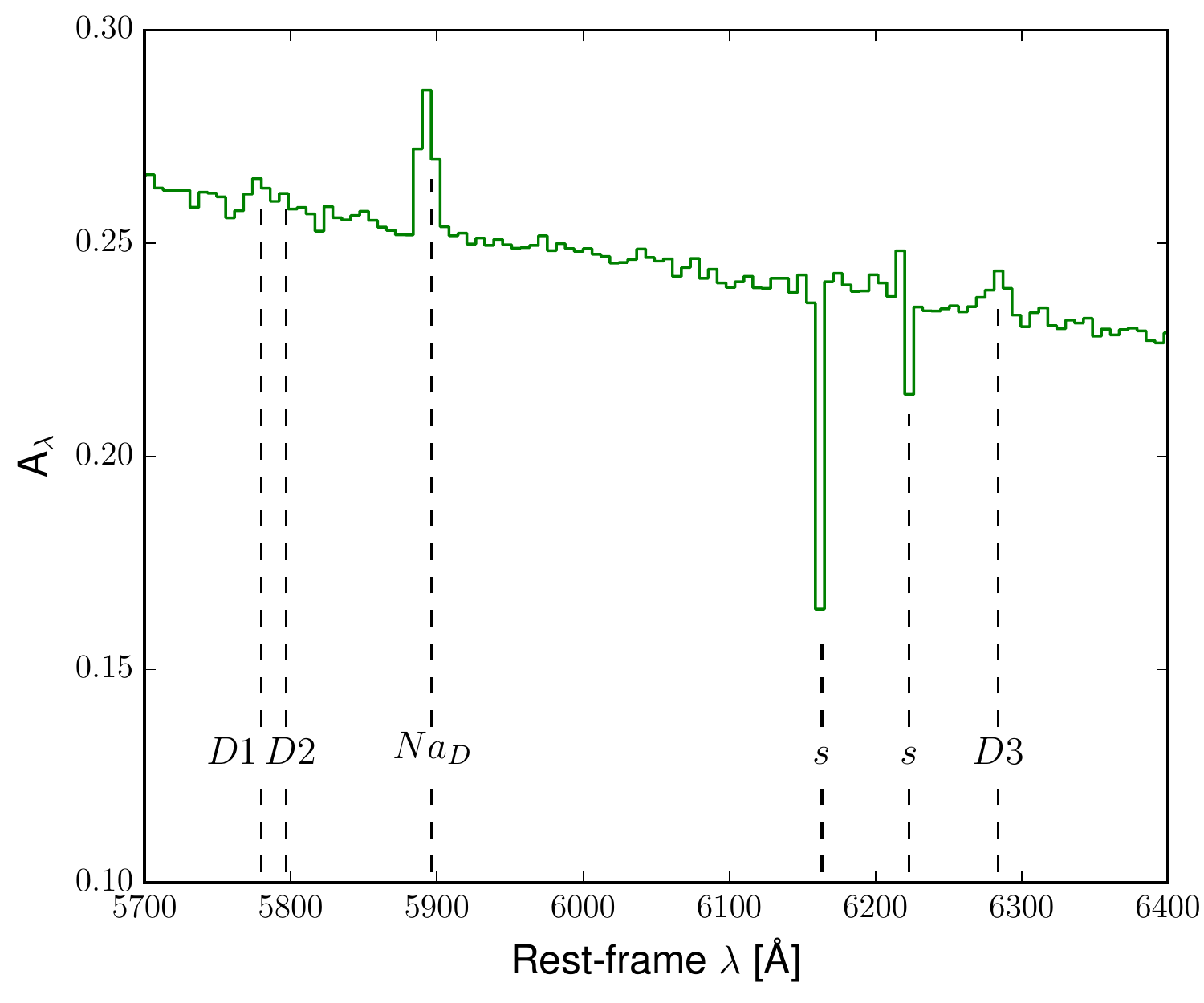}
    \caption{Zoom of the integrated attenuation curve of NGC 5626. Several spectral features are highlighted. Diffuse interstellar bands are detected in the Milky Way at 5780, 5797, and 6284 \AA. We indicated these wavelengths as D1 - D3, respectively. The sodium doublet (NaD) is not resolved, but clearly detected. Lines denotes with an `$s$' are residual sky-lines.}
    \label{fig:DIBs}
\end{figure}

We find interesting DIB candidates at rest-frame wavelengths of 5780, 5797, and 6284 \AA, which are also found along Milky Way sightlines \citep{Herbig1995}. Figure~\ref{fig:DIBs} shows a zoom of our attenuation curve, where these spectral features are indicated. The signal is strong enough to be detected in the integrated attenuation curve, but we found no significant detections in the attenuation curves of individual Voronoi bins.

Last, the sodium doublet is prominently visible in Fig.~\ref{fig:DIBs}, although the doublet is not spectrally resolved. We measure an $EW (Na_D)$ of $3.10$ \AA, which indicates significant absorption. In fact, \citet{Poznanski2012} found that the $E(B-V)$ vs. $EW (Na_D)$ relation saturates from $EW (Na_D) > 1$ for Galactic ISM clouds. The calibration offered by \citet{Baron2016} does extent to the high value we measure. However, their sample only contains AGN. Applying their relation, we find $E(B-V) = 0.31$ mag.

Due to a lack of $B$ band coverage (rest-frame wavelengths around $\sim 4450$ \AA\ are not covered by MUSE), we cannot directly compare this value. However, it should be higher than the rest-frame $E(4650-V)$ value of $0.07$ mag we measure. If we extrapolate the total attenuation curve to the $B$ band central wavelength, we obtain $A_B = 0.38$ mag and $E(B-V)=0.10$. For the extrapolation we use the best fit model for the global attenuation curve, discussed in Sect.~\ref{sec:modelling}. Both values are in line with each other, but significantly lower than the reddening derived from the \citet{Baron2016} relation. The lack of a proper calibration between $EW (Na_D)$ and dust reddening for normal galaxies certainly plays a role in this discrepancy. In addition, \citet{Poznanski2011} warns that measurements of the unresolved sodium doublet correlate poorly with reddening (in their case, for type Ia supernovae spectra). In contrast to the DIBs, we are able to spatially resolve the sodium doublet and investigate this further in Sect.~\ref{sec:NaDcorrelations}.

Often, extinction curves are characterized by their $R_V$ value,
\begin{equation}
R_\lambda = \frac{A_\lambda}{A_B - A_V},
\end{equation} 
which is 3.1 for the Milky Way \citep{Fitzpatrick1999}. Physically, smaller average dust grain sizes correspond to steeper extinction curves, and will have lower $R_V$ values. Vice versa, larger grains may be linked to higher $R_V$ values. However, one must be cautious into directly linking $R_V$ to dust grain properties because geometric and scattering effects heavily dillute this picture, as we discuss in Sect.~\ref{sec:discussion}. The estimate of $E(B-V)$ corresponds to a $V$ band total-to-selective attenuation $R_V$ of $2.8$. This compares to the value of $2.92 \pm 0.02$ found by \citet{Finkelman2008} for the same galaxy. Our estimate thus confirms a steeper attenuation curve than the Milky Way.

\subsubsection{Modelling} \label{sec:modelling}

The observed attenuation curve is often considered to be a probe for the intrinsic dust properties. In particular, the optical attenuation curve is usually linked to the average dust grain size \citep[see e.g.][]{Finkelman2010}. The underlying caveat here is that attenuation curves are treated as extinction curves. This means that the effects of scattering and the relative star-dust geometry are ignored.

Usually, both attenuation and extinction curves follow a power law in the optical, characterized by $R_V$. However, $R_V$ is limited to a very narrow range ($\Delta \lambda \approx 1200$ \AA), which limits its representation of the shape of the extinction curve. Furthermore, as mentioned above, the connection between $R_V$ and grain size is highly confounded when interpreting attenuation curves instead of extinction curves.

\begin{figure}
	\includegraphics[width=0.45\textwidth]{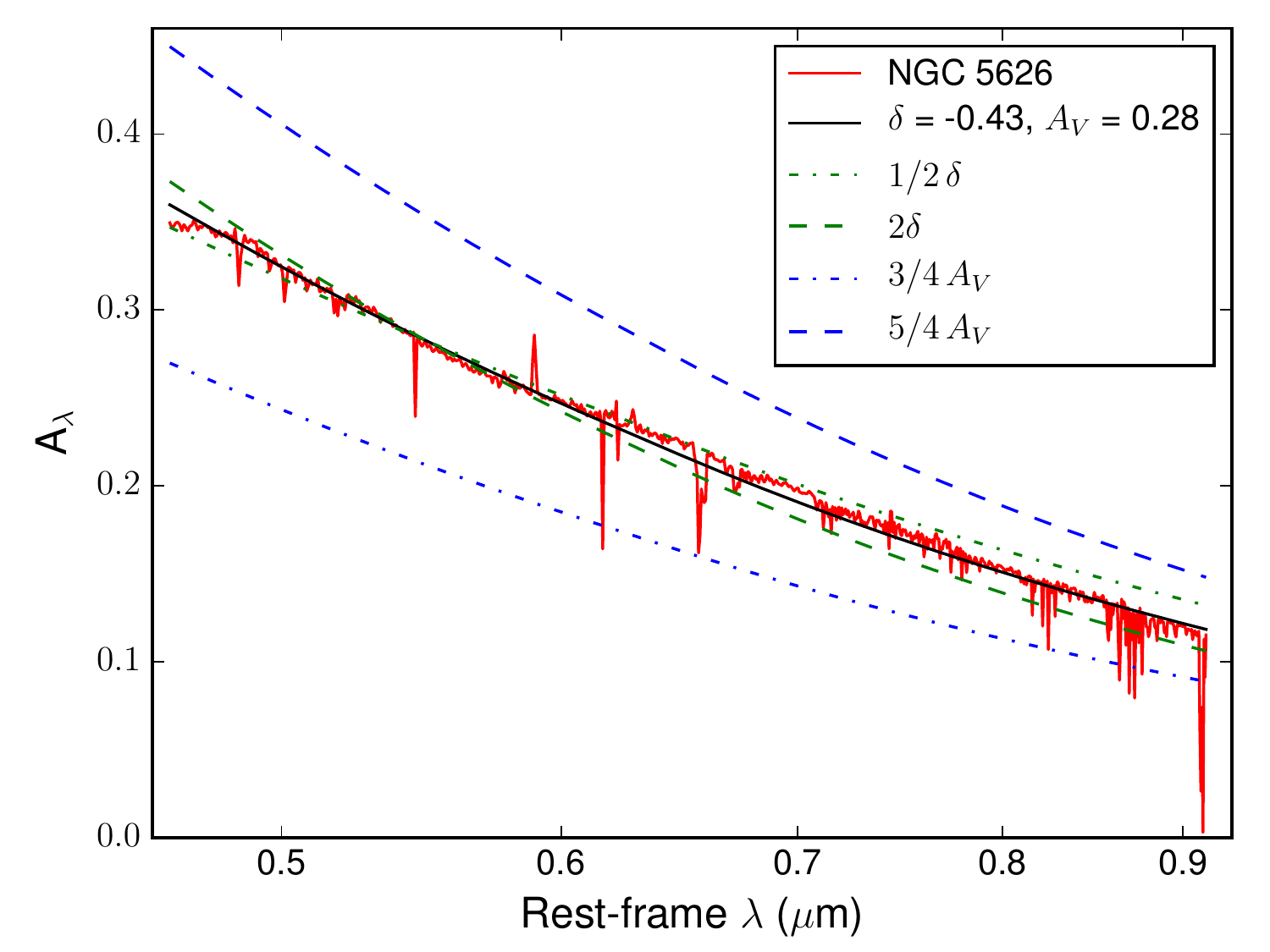}
    \caption{Models for the attenuation curve of NGC 5626. The black line is the best fit \textit{Calzetti-shape} model (see eq.~\eqref{eq:powerlaw}) to the observed attenuation curve. For reference, we also plot (in green) attenuation curve models for which $\delta$ is $50\%$ larger and smaller than the best-fit value (for the same value of $A_V$. In blue, the effect of a $25\%$ change in $A_V$ is shown, for a fixed $\delta$.}
    \label{fig:GlobalAfit}
\end{figure}

\begin{figure*}
	\includegraphics[width=\textwidth]{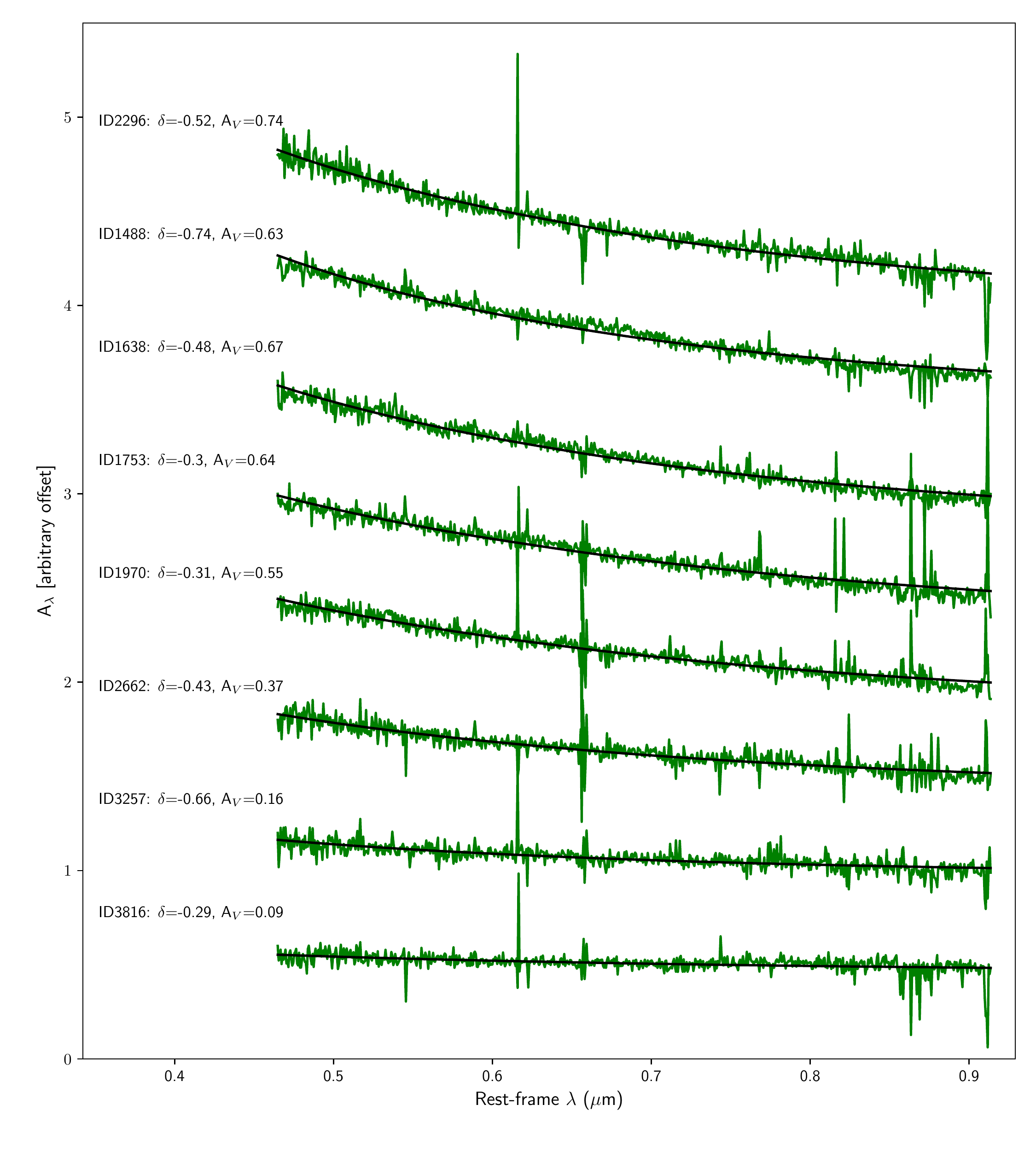}
    \caption{Attenuation curves for a representative selection of individual Voronoi bins (green). The black lines are the best fit models. Each curve was given an arbitrary offset for clarity. They are ranked by increasing $\Delta A$ (bottom to top) and the values for $\delta$ and $A_V$ are given for each curve. Less negative values of $\delta$ and small $A_V$ generally correspond to lower selective attenuation. However, there is no monotonic decrease in $\delta$ (or increase in $A_V$) when moving to the highest selective attenuation.}
    \label{fig:Afits}
\end{figure*}

For our measured attenuation curve, there are more than enough data points to constrain the shape, spanning a much wider wavelength range ($\Delta \lambda >4000$ \AA). We adopt the flexible and parametric attenuation curve model presented by \citet{Noll2009}. This is based on the general \textit{shape} of the \citet{Calzetti2000} attenuation curve $k$ (their eq.4):
\begin{equation} \label{eq:powerlaw}
A_\lambda = A_V \left( \frac{k(\lambda)}{k_V}\right) \left( \frac{\lambda}{0.55\, \mu\mathrm{m}}\right)^\delta .
\end{equation}
The two free parameters ($A_V$ and $\delta$) effectively provide a wide range in attenuation curve shapes. This formula is therefore relatively independent of the actual Calzetti normalisation and broadly applied to fit attenuation curves \citep{Buat2012, Kriek2013, Salmon2016, Seon2016}. We did not include the additional 2175 \AA \, bump in the model, as these authors did, because the observed wavelength range does not allow us to constrain the UV bump and would unnecessarily introduce new free parameters.

The best fitting model for the global attenuation curve yields $\delta= -0.43$ and $A_V=0.28$ and is shown in Fig.~\ref{fig:GlobalAfit}. We also plot model curves for higher and lower values of $\delta$, for the same $A_V$. This illustrates the effect of $\delta$ on the attenuation curve: higher values flatten the curve when $A_V$ is fixed. Additionally, the offset-effect of changing $A_V$ is shown for a fixed $\delta$. 

We also introduce a third quantity, to measure the selective attenuation over the full MUSE rest-frame wavelength range: 
\begin{equation}
\Delta A = A_{4655 \, \mu\mathrm{m}} - A_{9136 \, \mu\mathrm{m}}.
\end{equation}
$\Delta A = 0.24$ mag for the global model attenuation curve of NGC 5626. We will use this quantity as a natural replacement for $E(B-V)$ (which requires extrapolation) to investigate variations in selective attenuation on a local scale.
 
\subsection{Local variations of the attenuation curve} \label{sec:fitAcurves}

\begin{figure*}
	\includegraphics[width=\textwidth]{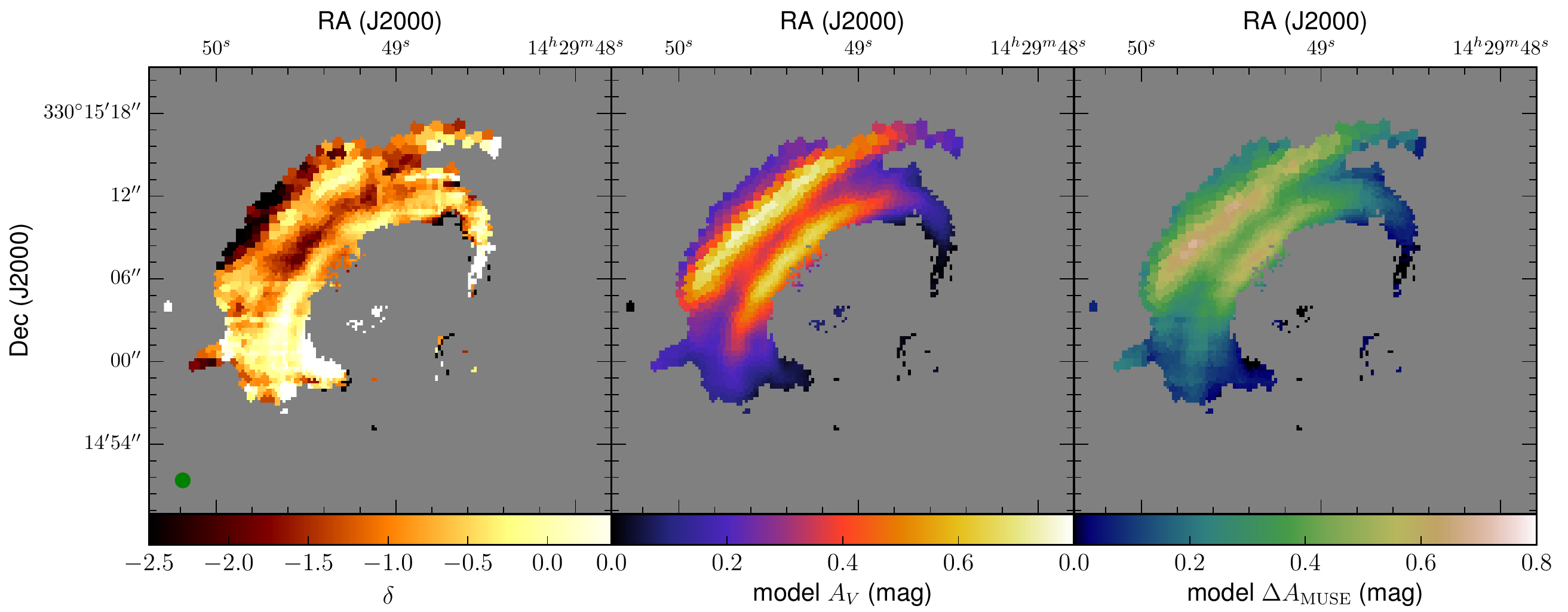}
    \caption{Parameter maps from the attenuation curves models (eq.~\eqref{eq:powerlaw}) for the Voronoi bins. Left: power-law index, $\delta$, which appears rather clumpy, but with a clear increase towards the middle of the dust lanes. Middle: model attenuation in the $V$ band, $A_V$, which is highest inside the dust lanes. Right: selective attenuation $\Delta A$ across the full $MUSE$ rest-frame wavelength range. High values correspond to a higher colour excess, which are strongly linked to regions of high attenuation (high $A_V$). The link with $\delta$ is more ambiguous, although the centre of the dust lanes usually correspond to steeper slopes.}
    \label{fig:AcurveFits}
\end{figure*}

\begin{figure}
	\includegraphics[width=0.45\textwidth]{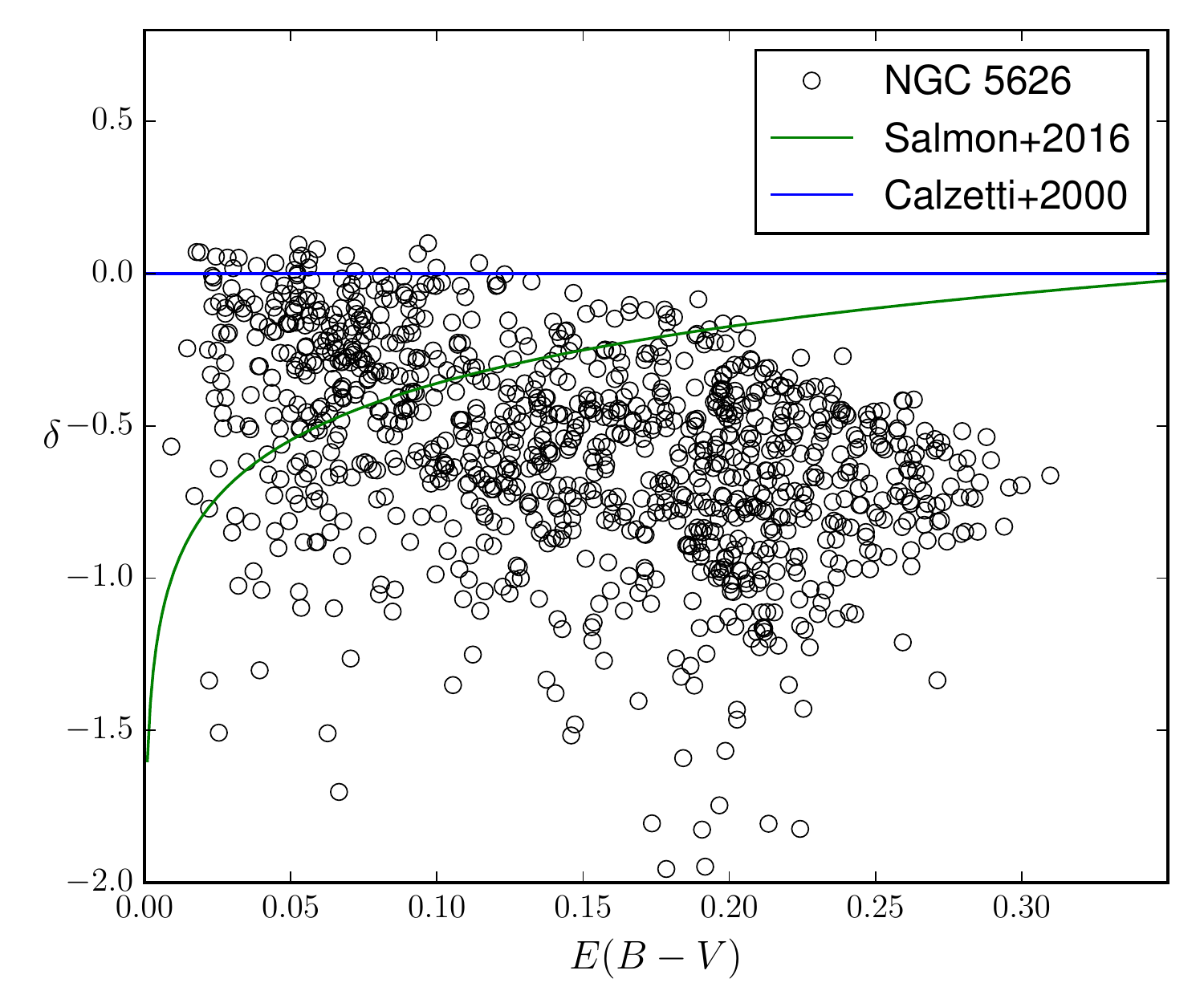}
    \caption{Estimated selective attenuation $E(B-V)$ vs power law index $\delta$ for attenuation curves in the dust lane. The blue line corresponds to the \citet{Calzetti2000} attenuation law for starburst galaxies. The green line is the fit from \citet{Salmon2016}, derived from dataset spanning a broader $E(B-V)$ range.}
    \label{fig:Ebv_delta}
\end{figure}

To compare the local attenuation curves in NGC 5626 in an objective way, we fit the attenuation curve in each Voronoi bin. Given the spread in size of the different Voronoi bins, this maps variations at the resolution of 0.2-2 arcsec, or a physical scale of $0.1-1$ kpc. Only regions with an average SNR above 2 in $A_\lambda$ were fitted, which corresponds mostly to the dust lane region.

We find significant variations in both strength and steepness of the attenuation curve. Several example fits are shown in Fig.~\ref{fig:Afits}. We show maps of $\delta$, $A_V$ and $\Delta A$ in Fig.~\ref{fig:AcurveFits}. There is a strong morphological correlation between $A_V$ and $\Delta A$, where a high colour excess occurs preferentially in the areas of higher $A_V$. 

The spatial distribution of $\delta$ is more ambiguous. We compare the values for $\delta$  and the selective attenuation of the curve for Voronoi bins in Fig.~\ref{fig:Ebv_delta}, following \citet{Salmon2016}. They fitted the attenuation curve of a sample of high-redshift galaxies with the same model as used here. They relate $\delta$ to the selective attenuation $E(B-V)$ in their figure 11. In general, they find that higher selective attenuation corresponds to higher values of $\delta$. However, the scatter is large and they even observe an opposite trend when looking only at galaxies with a strong evidence for SMC-like dust. The $E(B-V)$ parameter range for NGC 5626 is smaller than theirs, and does not fit in their general picture. However, it the trend does agree with the SMC-like attenuation curve that they observe for a subset of their sample.

\subsubsection{Converting attenuation to extinction}

In the classical view (i.e. interpreting the attenuation curves as extinction curves), the maps in Fig.~\ref{fig:AcurveFits} suggest that regions of higher reddening have smaller average grain sizes than regions in the outer part of the dust lane. However, this is based on the assumption of a linear relation between attenuation and extinction:
\begin{equation}
A_\lambda = 1.086 \tau_\lambda,
\end{equation}
which corresponds to an absorbing foreground screen dust geometry, ignoring effects due to scattering on dust grains.

It is of course more realistic that the dust in NGC 5626 is distributed in a ring or disk, embedded in the stellar distribution. The presence of several relaxed rings of ionized gas (as traced through H$\alpha$ emission) supports this assumption because dust and gas are nearly always spatially correlated \citep{Sarzi2006, Finkelman2010, Davis2011}. We see clear dust absorption North-East of the center, but little in the South-West. Given the inclined view we have of this system, this implies that the North-East side is closer to us. This implies that there are more stars \textit{behind} the dust ring or disk. Vice versa, there are more stars \textit{in front} of the dust in the South-West side. A foreground screen geometry is therefore not a realistic dust distribution.

\begin{figure}
	\includegraphics[width=0.5\textwidth]{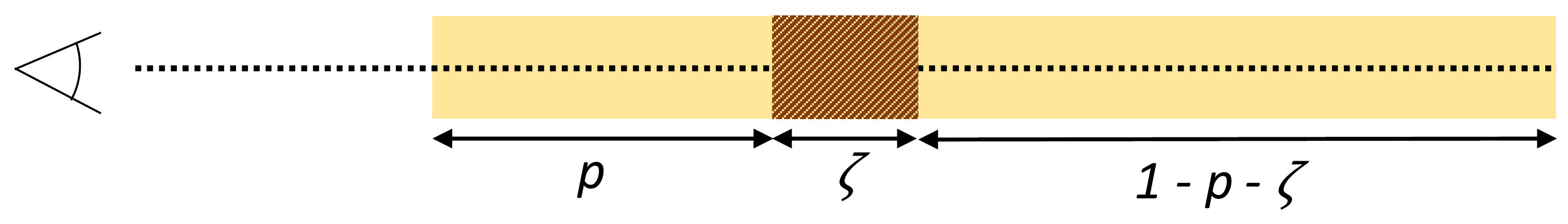}
    \caption{Schematic of the star-dust geometry described by eq.~\ref{eq:thicklayer}. A dust layer of thickness $\zeta$ is embedded in a stellar layer. A fraction of stars, $p$, lies in front of the dust and thus suffers no extinction.}
    \label{fig:ThickLayer}
\end{figure}

Instead, we consider an idealized toy model geometry, presented in Fig.~\ref{fig:ThickLayer}. Along each line of sight, there is a layer of stars, with a homogeneously mixed star-dust layer of a finite thickness $\zeta$ embedded. If $p$ is the fraction of stars in front of the dust layer, we can treat this fraction as unobscured. The fraction of stars behind the dust layer is then $1-p-\zeta$, and for these stars, the dust layer can be treated as a foreground screen. In total, we can model the absorption of the stellar flux along the line of sight as
\begin{equation} \label{eq:thicklayer}
F^\mathrm{obs}_\lambda = pF^0_\lambda + \zeta\frac{1-e^{-\tau_\lambda}}{\tau_\lambda} F^0_\lambda + (1-p-\zeta)e^{-\tau_\lambda} F^0_\lambda.
\end{equation}
Here, $F^{obs}_\lambda$ is the observed flux and $F^{0}_\lambda$ the total flux in the absence of dust, both for each wavelength $\lambda$. Equation~\ref{eq:thicklayer} describes the light extinction by a) stars in front of the dust (no extinction), b) stars inside the dust layer (uniformly mixed), c) stars behind the dust layer. Together they provide the attenuation of the star-dust system:
\begin{equation} \label{eq:thicklayerA}
\begin{split}
A_\lambda &= -2.5\log\left(\frac{F^\mathrm{obs}_\lambda}{F^0_\lambda}\right)  \\
          &= -2.5\log\left(p + \zeta\frac{1-e^{-\tau_\lambda}}{\tau_\lambda} + 				(1-p-\zeta)e^{-\tau_\lambda}\right) .
\end{split}
\end{equation}
This is a generalisation of the well-known `Sandwich model' for galaxies \citep{Disney1989, Boselli2003, Viaene2015}, with the addition that the dust layer can be moved along the line of sight (as described by $p$).

Any dust composition has its own extinction coefficient $\kappa_\lambda$ associated with it, which we assume to follow the same general shape as eq.~\eqref{eq:powerlaw}:	
\begin{equation} \label{eq:kappalaw}
\kappa_\lambda = \kappa_V \left( \frac{k(\lambda)}{k_V}\right) \left( \frac{\lambda}{0.55\, \mu\mathrm{m}}\right)^{\delta^\prime},
\end{equation}
with $\delta^\prime$ the power-law index for an extinction law. The total optical depth $\tau_\lambda$ then relates to the extinction coefficient through its integral of the dust density $\rho(s)$ along the line of sight $s$:
\begin{equation}
\tau_\lambda = \int_s \kappa_\lambda \rho (s) ds.
\end{equation}
For a particular 2D region on the sky (e.g. a Voronoi bin) we assume a constant $\kappa_\lambda$ along the line of sight, which allows us to write $\tau_\lambda = \Sigma_d \kappa_\lambda$, where $\Sigma_d$ is the dust surface density in that region on the sky.
We now investigate the effect of our geometric model on the observed attenuation curve for a given extinction law. 
For input relation for $\tau_\lambda$, we use $\delta^\prime=-0.66$ and $\kappa_V=0.73$ in arbitrary units. Furthermore, we fix $\Sigma_d$ to unity for all models in order to objectively compare them. Our choice for the numbers and (arbitrary units) for the input $\tau_\lambda$ are motivated such that for a simple screen geometry (where $A_V = 1.086 \tau_V$) we retrieve the attenuation curve with the highest colour excess.

\begin{figure}
	\includegraphics[width=0.5\textwidth]{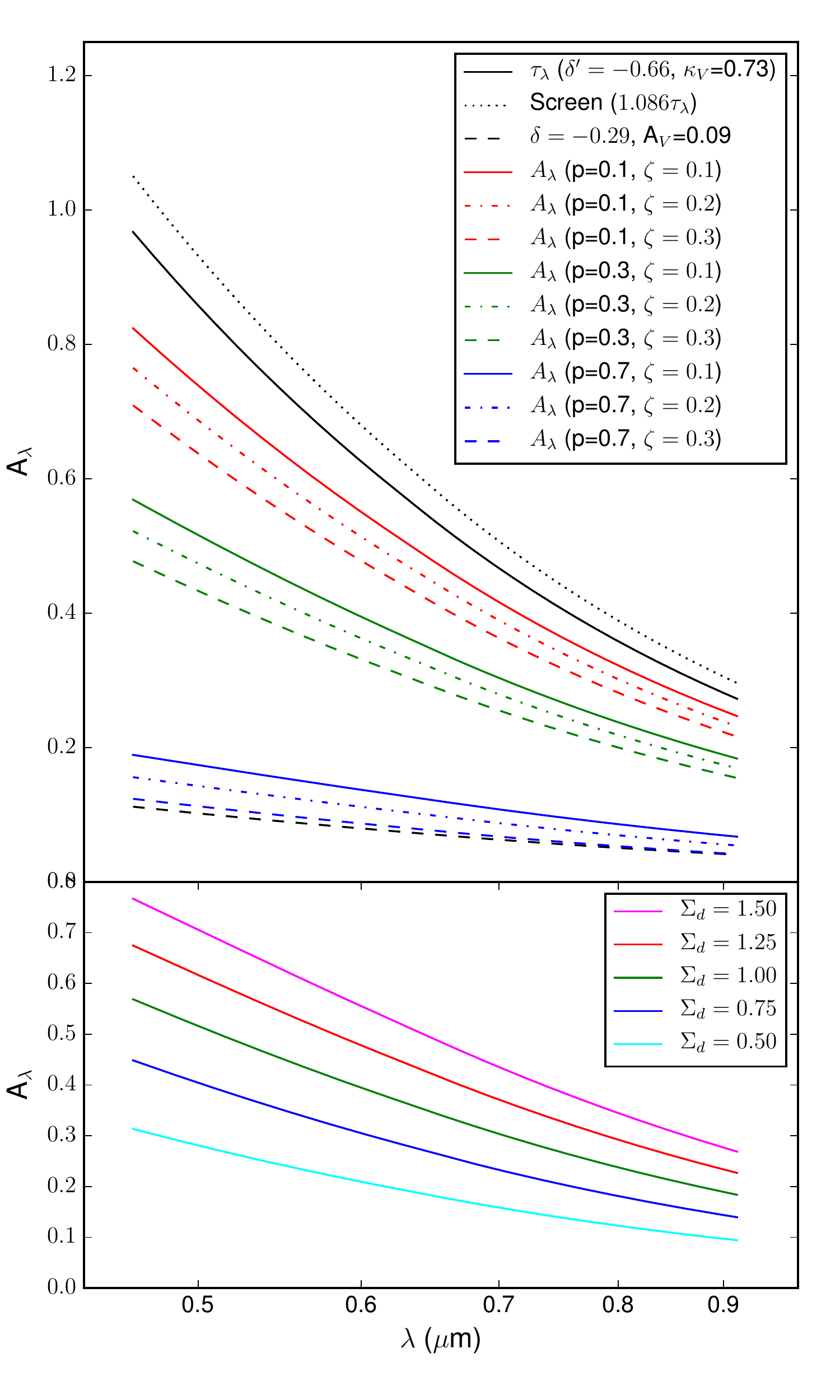}
    \caption{The effect of an embedded thick layer dust geometry on the reddening of starlight. Top panel: The black solid line is a \textit{Calzetti-shaped} curve used as an input extinction law (eq.~\eqref{eq:kappalaw}). The slope of the curve is unchanged for a dust screen geometry (black dotted line). The dust-starlight distribution corresponding to eq.~\eqref{eq:thicklayerA} lowers the extinction law to the red, green and blue lines. For reference, we also show the model fit for one of the observed attenuation curves with the lowest selective attenuation in NGC 5626 (black dashed line). Bottom panel: Effect of changing the dust mass (i.e. varying $\tau_V$) on the attenuation curve. A fixed intrinsic extinction law with $\delta^\prime=-0.66$ and $\kappa_V=0.73$ (according to  eq.~\eqref{eq:kappalaw}) was used. The geometrical parameters ($p=0.3$, $\zeta=0.1$) were also fixed, leaving the dust column density, $\Sigma_d$, as the only free parameter.}
    \label{fig:modelA_ThickLayer}
\end{figure}

Using eq.~\eqref{eq:thicklayerA}, we convert our input extinction law to an attenuation law. As one example, we take $p=0.3$, or $30 \%$ of the stars are in front of the dust lane, and $\zeta=0.1$, meaning the width of the dust lane is $10 \%$ of the width of the stellar distribution. Fig.~\ref{fig:modelA_ThickLayer} shows the corresponding attenuation curve (solid green line). The input extinction curve is significantly lower in selective attenuation under these geometrical conditions. When we fit again a power law model to the attenuation curve, we find that $\delta=-0.45$ and $A_V=0.45$. 

We also explore variations in $p$ and $\zeta$ in Fig.~\ref{fig:modelA_ThickLayer}. The major effect on the observed shape of the extinction curve can be ascribed to $p$. Varying $p$ from 0.1 to 0.7 can explain all combinations of $\delta$ and $A_V$ we measure for NGC 5626. As a reference, we show one of the power law fits with the lowest selective attenuation ($\delta=-0.29$ and $A_V=0.09$). As a second order effect, the thickness of the dust layer, $\zeta$ can also influence the shape. Thicker dust layers lower the selective attenuation.

A physical explanation of these effects lies in the fact that there are also stars in front of the dust lane. They are not attenuated, and thus bluer than the light that makes it through the dust lane. An observer simply records a mix of pure and reddened light. This mixture partly wipes out the wavelength-dependency of dust extinction. When interpreting attenuation curves as extinction curves, one basically assumes a foreground dust screen. In this situation, there are no stars in front of the dust, and so the shape of the extinction curve is not affected (see the dotted black line in Fig.~\ref{fig:modelA_ThickLayer}).

When increasing $p$, one essentially increases the amount of starlight in front of the dust lane, lowering the selective attenuation. The effect of increasing $\zeta$ (at fixed $p$) is similar. In essence, one takes stars that lie completely behind the dust layer and moves them inside the dust lane. While more stars radiate from inside the dust lane, their radiation has to pass through less dust than if $\zeta$ would be smaller. Therefore, the net result is a slightly lower colour excess.

\subsubsection{Discussion} \label{sec:discussion}

The above results highlight the degeneracy between geometry and intrinsic variations in the dust composition. However, one has to interpret the attenuation curve slope together with the absolute strength of the attenuation (e.g. $A_V$), and the global structure of the galaxy.

We return again to the maps in Fig.~\ref{fig:AcurveFits}. The signature of two dust lanes is clearly visible. We now assume that each dust lane is a disk or ring seen at some inclination. This is backed by the distribution of ionized gas as traced by H$\alpha$ (see Fig.~\ref{fig:HaMaps}). The regions of high attenuation thus correspond to the closest side of the rings. According to our simple model, $p$ will be relatively small. Indeed, we observe rather high $\Delta A$ Moving along a single dust ring towards the other side of the galaxy, we see the level of attenuation decrease. Gradually, more stars move in front of the dust along the line-of-sight. This corresponds to higher values for $p$. Hence, $\Delta A$ decreases.

While we can describe the variation of $\Delta A$ along a single dust lane, it is harder to explain variations in the direction perpendicular to the dust lane. Moving outward from the center along the minor axis, $p$ should decrease monotonically as fewer stars lie in front of the dust. Yet we observe lower $\Delta A$ values in between the two dust rings. At the same time, we find lower values of $A_V$. In our interpretation of the model described by eq.~\eqref{eq:thicklayerA}, we started from a fixed function for $\tau_\lambda$ (i.e. a fixed dust model). This is reasonable for an axisymmetric ring of dust, but isn't necessarily true when moving from small to large galactocentric distances. For example, \citet{Smith2012b} found radial variations in mass and dust properties in the Andromeda galaxy (M31).

A change in dust mass along the line-of-sight implies implies changing $\Sigma_d$, while leaving $\kappa_\lambda$ fixed. This induces a multiplication of the extinction law with a non-unity factor, effectively changing the optical depth curve $\tau_\lambda$. We explore the corresponding attenuation curves in in the bottom panel of Fig.~\ref{fig:modelA_ThickLayer}. The dust column density shows a strong influence both the strength and the gradient of the attenuation curve. In fact, a decrease in $\Sigma_d$ by a factor of 2.2 can already  change in $A_V$ from $0.75$ to $0.40$, representative values for the outer dust lane, and the region in between the two dust lanes, respectively.

Our results are consistent with uniform dust properties across the galaxy. However, the dust distribution in NGC 5626 may not span the full extent of the explored parameter ranges in geometry ($p$, $\zeta$) and column density ($\Sigma_d$), which would leave the door open for intrinsic variations in the dust properties. Nevertheless, the explored parameter space is narrow and we consider all values probable. The main lesson learned here is that the degeneracy between distribution, mass, and intrinsic properties is strong. It places strong limits on the interpretation and comparison of measured attenuation curves across galaxies, or within a galaxy.

As a final remark, we underline that scattering of light on dust grains is not included in this analytical model. \citet{Baes2001} showed, in a series of radiative transfer simulations, that scattering tends to flatten the attenuation curve slope. However, they warn that the effects can vary significantly depending on both geometry and optical depth. \citet{Scicluna2015} also performed a series of radiative transfer simulations for different dust distributions. They found that scattering is predominantly affecting the attenuation curve when the line of sight contains deeply embedded structures such as AGN. On kilo-parsec scales and for more diffuse dust distributions, the effects are smaller. We warn that scattering will further dilute the interpretation of attenuation curves in the light of dust grain variations. This will happen for all lines of sight, both along the dust lanes, and in the direction perpendicular to them.

Incident light has a certain probability to be scattered or absorbed, depending on the optical properties of the dust grains. It furthermore couples multiple sightlines. Precisely due to this stochastic process, it is not possible to mimic scattering effects with pure analytical models such as eq.~\eqref{eq:thicklayerA}. Intermediate solutions have been explored in the past, such as different optical depths for homogeneously mixed dust and foreground dust \citep{Natta1984, Calzetti1994}. This method partially folds scattering effects into  the optical depth parametrisation. However, with the recent increase in computing power, the most straightforward way to include scattering is through radiative transfer simulations. They embrace the 3D, non-local nature of the problem by combining geometry with both absorption and scattering. We plan such simulations in future work.

\subsection{Line tracers of dust reddening}

The sensitive high-resolution IFU power of MUSE not only allows the analysis presented above. One can simultaneously look at line tracers of dust reddening. The sodium doublet and the Balmer decrement are two primary examples of spectral features that are linked to reddening by dust, and are relatively easy to measure.

\subsubsection{Sodium doublet} \label{sec:NaDcorrelations}

\begin{figure}
	\includegraphics[width=0.45\textwidth]{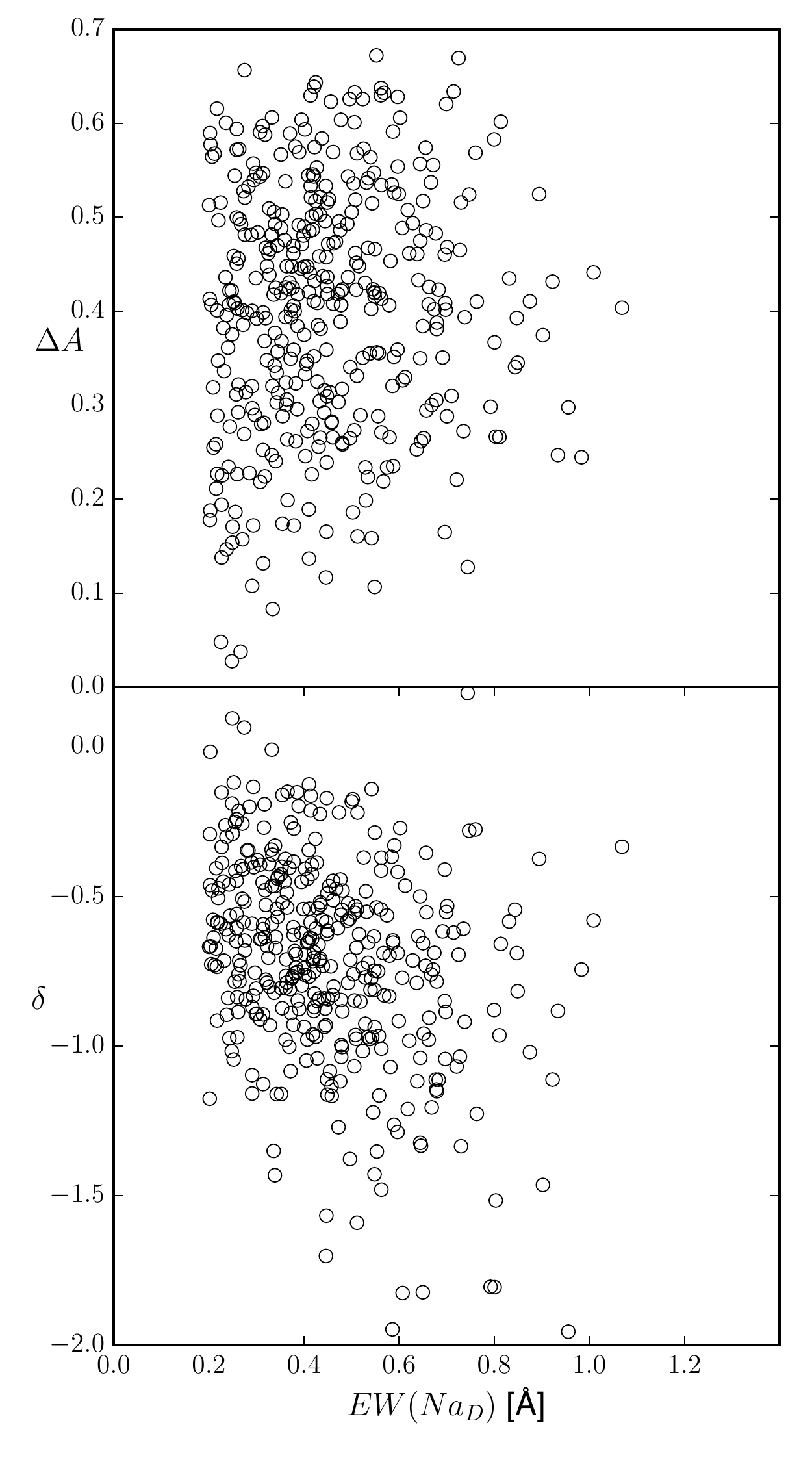}
    \caption{Tracers of dust reddening as a function of Na$_{\text{D}}$ measurements for the individual Voronoi bins. Top panel shows the $\Delta A$ metric, which reflects the selective attenuation for the full MUSE rest-frame wavelength range. Bottom panel shows the power-law index $\delta$. The correlation in both panels is very weak. On the physical scales probed here ($0.1-1$ kpc), the strength of the sodium line does not relate to dust reddening.}
    \label{fig:NaDcorrelations}
\end{figure}

The association between the Na$_\text{D}$ doublet absorption line and dust reddening has been made for Galactic lines of sight \citep{Poznanski2012}, and for AGN spectra \citep{Baron2016}. Both studies attribute the reddening to cold ISM clouds, which contain diffuse dust. Given the low ionisation potential of sodium ($5.14$ eV), this element is also expected to be present in these cold environments. It remains uncertain, however, how both components are related, on a spatially resolved scale, in normal galaxies. Using our GandALF model for the MUSE datacube of NGC 5626, we are able to measure the spectrally unresolved NaD line in Voronoi bins associated with the dust lane.
 
In Fig.~\ref{fig:NaDcorrelations}, we plot the correlation of EW(Na$_{\text{D}}$) versus $\Delta A$ and $\delta$, two parameters linked to dust reddening. Note again that there is no $B$ band data available to actually measure quantities like R$_V$ or $E(B-V)$. Perhaps surprisingly, we find no evidence of a correlation between EW(Na$_{\text{D}}$) and either $\Delta A$ or $\delta$.

It appears that, on the spatial scales we reach in NGC 5626 ($0.1-1$ kpc), there is no connection between the sodium doublet and dust reddening. On the other hand, the Na$_\text{D}$ doublet is clearly detected when averaging the flux of dust-affected regions (see again Fig.~\ref{fig:att_norm}, bottom panel). Sodium also occurs in cooler stellar atmospheres and indeed we do detect it in the dust-free areas of NGC 5626. \citet{Poznanski2012} used spectra of background QSOs to investigate EW(Na$_{\text{D}}$) vs. reddening in the Milky Way. With their technique they were effectively probing single cold clouds, which we cannot. 
In addition, as already noted in Sect.~\ref{sec:globalAtt}, measurements of the unresolved sodium doublet show a weaker correlation with reddening than resolved line measurements do \citep{Poznanski2011}.

\subsubsection{Balmer decrement}

\begin{figure}
	\includegraphics[width=0.45\textwidth]{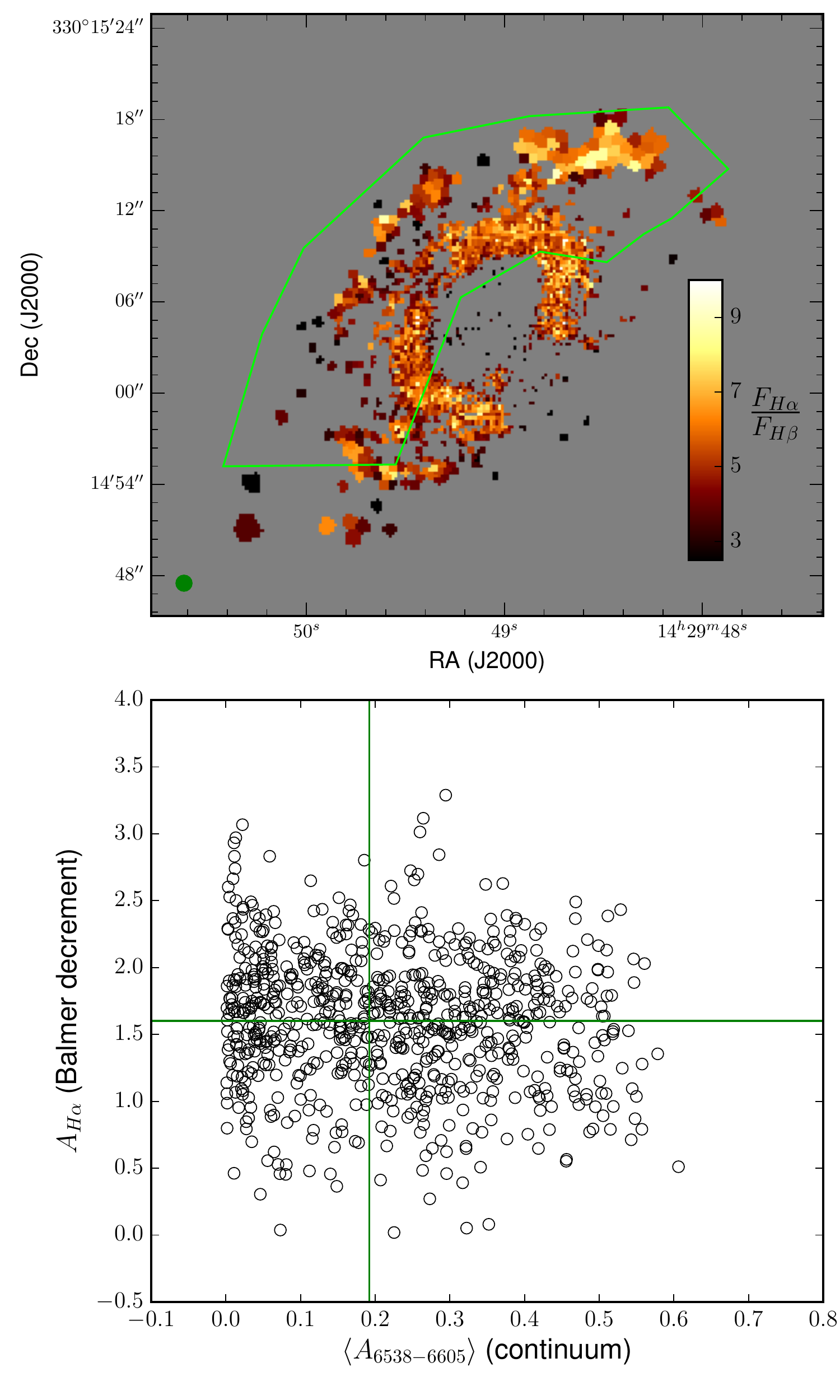}
    \caption{Balmer decrement in NGC 5626. Top: Map of the $F_{\text{H}\alpha}/F_{\text{H}\beta}$ ratio, where it is detected. Higher values point to more extinction. No extinction corresponds to the theoretical ratio of $2.85$. Bottom: Continuum attenuation near the observed H$\alpha$ line v.s. the H$\alpha$ line attenuation derived from the Balmer decrement. The green lines mark the average in both parameters, yielding an average $A_{\mathrm{H}\alpha} / A_\mathrm{cont}$ ratio of $8.3 \pm 7.1$.}
    \label{fig:balmer}
\end{figure}

The line flux ratio between H$\alpha$ and H$\beta$ is often used to de-redden the H$\alpha$ flux \citep[see e.g.][and references therein]{Gilbank2010, Groves2012, Dominguez2013}. The Balmer decrement is shown for NGC 5626 in Fig.~\ref{fig:balmer}. Its determination is mainly driven by whether H$\beta$ is detected in a Voronoi bin or not. While it is difficult to disentangle differences in strength, the main morphology is clear. A clear central ring is visible, similar to the one observed in the H$\alpha$ map (Fig.~\ref{fig:HaMaps}). There are also three segments visible towards the South-East, East, and North-West, respectively. Their position suggests that they are part of an second ring or a spiral structure.

Given the Balmer decrement (i.e. the flux ratio $F_{H\alpha}/F_{H\beta}$), we can estimate the attenuation of the H$\alpha$ emission (see for example equation 3 in \citealt{Gilbank2010}):
\begin{equation}
A_{\text{H}\alpha} = \frac{2.5}{\kappa_{\text{H}\beta}/\kappa_{\text{H}\alpha}-1} \log\left(\frac{1}{2.85} \frac{F_{\text{H}\alpha}}{F_{\text{H}\beta}} \right),
\end{equation}
where $\kappa_{\text{H}\beta}$ and $\kappa_{\text{H}\alpha}$ are the extinction coefficients at the rest-frame wavelengths of H$\beta$ and H$\alpha$, respectively. We use the values from the THEMIS extinction curve, calibrated for MW dust \citep{Ysard2015, Jones2017}. This formula assumes case-B recombination for the intrinsic Balmer decrement of 2.85 \citep{Osterbrock1989}.

In Fig.~\ref{fig:balmer}, we compare the attenuation derived from the Balmer decrement to the attenuation in the continuum, $\langle A_{6538-6605}\rangle$. The latter is derived from our MGE fitting effort. We take, for each Voronoi bin, the average of the attenuation at $6538$ \AA\ and at $6605$ \AA. These wavelengths correspond to continuum emission just shortwards and longwards of the observed H$\alpha$+$NI$ line. Remarkably, we find no correlation between the continuum and the line attenuation at the H$\alpha$ wavelength. In fact, the line attenuation appears to be constant for any value of the continuum attenuation. We also plot the mean value of $A_{\text{H}\alpha}$ for the entire sample, which is $1.60$, with a standard deviation of $0.55$. The mean value for $\langle A_{6538-6605}\rangle$ is $0.19$ with a standard deviation of $0.15$, yielding an average $A_{\mathrm{H}\alpha} / A_\mathrm{cont}$ ratio of $8.3 \pm 7.1$.

The complete insensitivity of the line attenuation for changing continuum attenuation points towards different attenuation mechanisms. As attenuation is the convolution of geometry and extinction, this suggests that both forms of attenuation are either geometrically decoupled, or have different extinction laws. Both phenomena can be interpreted as line attenuation occurring predominantly in star forming clouds. \citet{Calzetti1994} already described the phenomenon of selective extinction, where line emission is obscured by both by clouds and the diffuse ISM, while continuum emission only suffers extinction from the diffuse dust component \citep[see also][]{Charlot2000, Calzetti2001, Wild2011}. As expected, $A_{\text{H}\alpha}$ is consistently higher than $\langle A_{6538-6605}\rangle$. In addition to the geometrical effect, cloud environments are denser than the general ISM, which could influence grain size distributions (and so the extinction law).

It is in fact likely that a combination of geometry and intrinsic extinction provokes the decoupling between line and continuum attenuation. This limits the use of the Balmer decrement as a general tracer of dust attenuation in galaxies \citep[e.g.][]{Dominguez2013, Reddy2015, Nelson2016, Battisti2016}. For a set of nearby galaxies, \citet{Kreckel2013} found a positive trend between dust column density (as traced by dust emission), and $A_{V}$ derived using the Balmer decrement and a Calzetti-attenuation law. However, the correlation shows significant scatter between different galaxies, which they attribute to resolution effects. Our results add to their warning for the use of Balmer decrement as a global recipe for de-reddening stellar spectra or SEDs.

\section{Summary \& conclusions} \label{sec:conclusions}

We have investigated the dust and ionized gas in NGC 5626 using IFU spectrographic data from the MUSE science verification phase. This was an effort to demonstrate the capabilities of MUSE to study the ISM in early-type galaxies.
We measured the optical attenuation curve for both the galaxy as a whole, and for each spatial resolution element. To our knowledge, this is the first truly spectral-resolution (6.25 \AA) measurement of an optical attenuation curve of an \textit{individual} galaxy. We find that:

\begin{itemize}
\item The integrated attenuation curve of NGC 5626 follows a smooth trend, with several sharp features. There is a strong decrease in attenuation at the position of emission lines in the spectrum (e.g. H$\alpha$ and H$\beta$). This result could be driven by the clumpy nature of the line emission, which is not captured in the modeling of the continuum emission. Vice versa, the attenuation is significantly higher for the sodium absorption line doublet.

\item We detect three candidates for diffuse interstellar bands in the integrated attenuation curve. Unfortunately, the signal is too low to map them spatially.

\item We plotted the strength (equivalent width) of the sodium absorption doublet against the reddening by dust (traced by the selective attenuation). While previous studies found a positive correlation for diffuse Milky Way dust, there is no correlation for NGC 5626 at our spatial resolution ($0.1-1$ kpc). 

\item At the same spatial scales, the Balmer decrement yields H$\alpha$ attenuation values that are completely independent from the continuum attenuation at H$\alpha$ wavelenghts. The decoupling of line and continuum attenuation suggests a combination of geometric decoupling and different extinction laws for both components. This preaches caution on the use of the Balmer decrement to de-redden stellar spectra.
\end{itemize}

The attenuation curve for each resolution element is fitted with a power law to derive the slope and $V$ band attenuation level. A simple `embedded thick layer' model is presented to parametrise the star-dust distribution along the line of sight and to interpret the variation of the observed attenuation curve shapes.

\begin{itemize}
\item We find that the amount and distribution of dust along the line-of-sight is highly degenerate with any variation in the intrinsic extinction law. While spatial variations in the dust composition are certainly possible, the retrieved attenuation maps of NGC 5626 are also consistent with a single dust composition across the galaxy. Adding up the evidence from our attenuation maps, the H$\alpha$ line strength and velocity distribution, the balmer decrement, and our geometrical analysis. We conclude that the ISM in this galaxy resides in a regular and well-settled disk.

\end{itemize}

The new capabilities of IFU observations with MUSE bring us closer to 3D mapping of the ISM in galaxies. Despite the increased resolution it remains remarkably hard to make the distinction between geometric effects and intrinsic dust property variations. One should therefore be cautious to correct UV/optical/NIR images for the effects of dust before analysing the stellar properties. This pilot study should be continued for similar systems to assess whether a single attenuation curve is a good approximation to correct spatially resolved images of galaxies for the effects of dust. In addition, we advocate the use of radiative transfer simulations to mimic the effects of multiple scattering, and 3D star-dust distributions. We will construct such models in future work.

\section*{Acknowledgements}

We wish to thank the anonymous referee for the positive feedback which has improved the paper significantly. We also thank Borislav Nedelchev for stimulating discussions regarding this topic. \\
This research is funded by the BOF of Ghent University. SV acknowledges the support of the FWO mobility grant. \\
Based on observations collected at the European Organisation for Astronomical Research in the Southern Hemisphere under ESO programme 60.A-9337(A) as part of the MUSE Science Verification phase. 




\bibliographystyle{mnras}
\bibliography{references} 

\begin{thebibliography}{}
\makeatletter
\relax
\def\mn@urlcharsother{\let\do\@makeother \do\$\do\&\do\#\do\^\do\_\do\%\do\~}
\def\mn@doi{\begingroup\mn@urlcharsother \@ifnextchar [ {\mn@doi@}
  {\mn@doi@[]}}
\def\mn@doi@[#1]#2{\def\@tempa{#1}\ifx\@tempa\@empty \href
  {http://dx.doi.org/#2} {doi:#2}\else \href {http://dx.doi.org/#2} {#1}\fi
  \endgroup}
\def\mn@eprint#1#2{\mn@eprint@#1:#2::\@nil}
\def\mn@eprint@arXiv#1{\href {http://arxiv.org/abs/#1} {{\tt arXiv:#1}}}
\def\mn@eprint@dblp#1{\href {http://dblp.uni-trier.de/rec/bibtex/#1.xml}
  {dblp:#1}}
\def\mn@eprint@#1:#2:#3:#4\@nil{\def\@tempa {#1}\def\@tempb {#2}\def\@tempc
  {#3}\ifx \@tempc \@empty \let \@tempc \@tempb \let \@tempb \@tempa \fi \ifx
  \@tempb \@empty \def\@tempb {arXiv}\fi \@ifundefined
  {mn@eprint@\@tempb}{\@tempb:\@tempc}{\expandafter \expandafter \csname
  mn@eprint@\@tempb\endcsname \expandafter{\@tempc}}}

\bibitem[\protect\citeauthoryear{{Bacon} et~al.,}{{Bacon} et~al.}{2010}]{MUSE}
{Bacon} R.,  et~al., 2010, in Ground-based and Airborne Instrumentation for
  Astronomy III. p. 773508, \mn@doi{10.1117/12.856027}

\bibitem[\protect\citeauthoryear{{Baes} \& {Dejonghe}}{{Baes} \&
  {Dejonghe}}{2001}]{Baes2001}
{Baes} M.,  {Dejonghe} H.,  2001, \mn@doi [\mnras]
  {10.1046/j.1365-8711.2001.04626.x}, \href
  {http://adsabs.harvard.edu/abs/2001MNRAS.326..733B} {326, 733}

\bibitem[\protect\citeauthoryear{{Baron}, {Poznanski}, {Watson}, {Yao}  \&
  {Prochaska}}{{Baron} et~al.}{2015}]{Baron2015}
{Baron} D.,  {Poznanski} D.,  {Watson} D.,  {Yao} Y.,   {Prochaska} J.~X.,
  2015, \mn@doi [\mnras] {10.1093/mnras/stu2448}, \href
  {http://adsabs.harvard.edu/abs/2015MNRAS.447..545B} {447, 545}

\bibitem[\protect\citeauthoryear{{Baron}, {Stern}, {Poznanski}  \&
  {Netzer}}{{Baron} et~al.}{2016}]{Baron2016}
{Baron} D.,  {Stern} J.,  {Poznanski} D.,   {Netzer} H.,  2016, \mn@doi [\apj]
  {10.3847/0004-637X/832/1/8}, \href
  {http://adsabs.harvard.edu/abs/2016ApJ...832....8B} {832, 8}

\bibitem[\protect\citeauthoryear{{Battisti}, {Calzetti}  \& {Chary}}{{Battisti}
  et~al.}{2016}]{Battisti2016}
{Battisti} A.~J.,  {Calzetti} D.,   {Chary} R.-R.,  2016, \mn@doi [\apj]
  {10.3847/0004-637X/818/1/13}, \href
  {http://adsabs.harvard.edu/abs/2016ApJ...818...13B} {818, 13}

\bibitem[\protect\citeauthoryear{{Battisti}, {Calzetti}  \& {Chary}}{{Battisti}
  et~al.}{2017}]{Battisti2017}
{Battisti} A.~J.,  {Calzetti} D.,   {Chary} R.-R.,  2017, \mn@doi [\apj]
  {10.3847/1538-4357/aa6fb2}, \href
  {http://adsabs.harvard.edu/abs/2017ApJ...840..109B} {840, 109}

\bibitem[\protect\citeauthoryear{{Bertola} \& {Galletta}}{{Bertola} \&
  {Galletta}}{1978}]{Bertola1978}
{Bertola} F.,  {Galletta} G.,  1978, \mn@doi [\apjl] {10.1086/182844}, \href
  {http://adsabs.harvard.edu/abs/1978ApJ...226L.115B} {226, L115}

\bibitem[\protect\citeauthoryear{{Boselli}, {Gavazzi}  \& {Sanvito}}{{Boselli}
  et~al.}{2003}]{Boselli2003}
{Boselli} A.,  {Gavazzi} G.,   {Sanvito} G.,  2003, \mn@doi [\aap]
  {10.1051/0004-6361:20030219}, \href
  {http://adsabs.harvard.edu/abs/2003A%26A...402...37B} {402, 37}

\bibitem[\protect\citeauthoryear{{Boselli} et~al.,}{{Boselli}
  et~al.}{2010}]{Boselli2010}
{Boselli} A.,  et~al., 2010, \mn@doi [\pasp] {10.1086/651535}, \href
  {http://adsabs.harvard.edu/abs/2010PASP..122..261B} {122, 261}

\bibitem[\protect\citeauthoryear{{Brosch}, {Greenberg}  \& {Grosbol}}{{Brosch}
  et~al.}{1985}]{Brosch1985}
{Brosch} N.,  {Greenberg} J.~M.,   {Grosbol} P.~J.,  1985, \aap, \href
  {http://adsabs.harvard.edu/abs/1985A%26A...143..399B} {143, 399}

\bibitem[\protect\citeauthoryear{{Brosch}, {Almoznino}, {Grosbol}  \&
  {Greenberg}}{{Brosch} et~al.}{1990}]{Brosch1990}
{Brosch} N.,  {Almoznino} E.,  {Grosbol} P.,   {Greenberg} J.~M.,  1990, \aap,
  \href {http://adsabs.harvard.edu/abs/1990A%26A...233..341B} {233, 341}

\bibitem[\protect\citeauthoryear{{Buat} et~al.,}{{Buat}
  et~al.}{2012}]{Buat2012}
{Buat} V.,  et~al., 2012, \mn@doi [\aap] {10.1051/0004-6361/201219405}, \href
  {http://adsabs.harvard.edu/abs/2012A%26A...545A.141B} {545, A141}

\bibitem[\protect\citeauthoryear{{Bundy} et~al.,}{{Bundy} et~al.}{2015}]{MaNGA}
{Bundy} K.,  et~al., 2015, \mn@doi [\apj] {10.1088/0004-637X/798/1/7}, \href
  {http://adsabs.harvard.edu/abs/2015ApJ...798....7B} {798, 7}

\bibitem[\protect\citeauthoryear{{Byun}, {Freeman}  \& {Kylafis}}{{Byun}
  et~al.}{1994}]{Byun1994}
{Byun} Y.~I.,  {Freeman} K.~C.,   {Kylafis} N.~D.,  1994, \mn@doi [\apj]
  {10.1086/174553}, \href {http://adsabs.harvard.edu/abs/1994ApJ...432..114B}
  {432, 114}

\bibitem[\protect\citeauthoryear{{Calzetti}}{{Calzetti}}{2001}]{Calzetti2001}
{Calzetti} D.,  2001, \mn@doi [\pasp] {10.1086/324269}, \href
  {http://adsabs.harvard.edu/abs/2001PASP..113.1449C} {113, 1449}

\bibitem[\protect\citeauthoryear{{Calzetti}, {Kinney}  \&
  {Storchi-Bergmann}}{{Calzetti} et~al.}{1994}]{Calzetti1994}
{Calzetti} D.,  {Kinney} A.~L.,   {Storchi-Bergmann} T.,  1994, \mn@doi [\apj]
  {10.1086/174346}, \href {http://adsabs.harvard.edu/abs/1994ApJ...429..582C}
  {429, 582}

\bibitem[\protect\citeauthoryear{{Calzetti}, {Armus}, {Bohlin}, {Kinney},
  {Koornneef}  \& {Storchi-Bergmann}}{{Calzetti} et~al.}{2000}]{Calzetti2000}
{Calzetti} D.,  {Armus} L.,  {Bohlin} R.~C.,  {Kinney} A.~L.,  {Koornneef} J.,
   {Storchi-Bergmann} T.,  2000, \mn@doi [\apj] {10.1086/308692}, \href
  {http://adsabs.harvard.edu/abs/2000ApJ...533..682C} {533, 682}

\bibitem[\protect\citeauthoryear{{Campbell}, {Holz}, {Gerlich}  \&
  {Maier}}{{Campbell} et~al.}{2015}]{Campbell2015}
{Campbell} E.~K.,  {Holz} M.,  {Gerlich} D.,   {Maier} J.~P.,  2015, \mn@doi
  [\nat] {10.1038/nature14566}, \href
  {http://adsabs.harvard.edu/abs/2015Natur.523..322C} {523, 322}

\bibitem[\protect\citeauthoryear{{Cappellari}}{{Cappellari}}{2002}]{Cappellari2002}
{Cappellari} M.,  2002, \mn@doi [\mnras] {10.1046/j.1365-8711.2002.05412.x},
  \href {http://adsabs.harvard.edu/abs/2002MNRAS.333..400C} {333, 400}

\bibitem[\protect\citeauthoryear{{Cappellari}}{{Cappellari}}{2016}]{Cappellari2016}
{Cappellari} M.,  2016, \mn@doi [\araa] {10.1146/annurev-astro-082214-122432},
  \href {http://adsabs.harvard.edu/abs/2016ARA%26A..54..597C} {54, 597}

\bibitem[\protect\citeauthoryear{{Charlot} \& {Fall}}{{Charlot} \&
  {Fall}}{2000}]{Charlot2000}
{Charlot} S.,  {Fall} S.~M.,  2000, \mn@doi [\apj] {10.1086/309250}, \href
  {http://adsabs.harvard.edu/abs/2000ApJ...539..718C} {539, 718}

\bibitem[\protect\citeauthoryear{{Ciesla} et~al.,}{{Ciesla}
  et~al.}{2016}]{Ciesla2016}
{Ciesla} L.,  et~al., 2016, \mn@doi [\aap] {10.1051/0004-6361/201527107}, \href
  {http://adsabs.harvard.edu/abs/2016A%26A...585A..43C} {585, A43}

\bibitem[\protect\citeauthoryear{{Clayton}, {Gordon}, {Bianchi}, {Massa},
  {Fitzpatrick}, {Bohlin}  \& {Wolff}}{{Clayton} et~al.}{2015}]{Clayton2015}
{Clayton} G.~C.,  {Gordon} K.~D.,  {Bianchi} L.~C.,  {Massa} D.~L.,
  {Fitzpatrick} E.~L.,  {Bohlin} R.~C.,   {Wolff} M.~J.,  2015, \mn@doi [\apj]
  {10.1088/0004-637X/815/1/14}, \href
  {http://adsabs.harvard.edu/abs/2015ApJ...815...14C} {815, 14}

\bibitem[\protect\citeauthoryear{{Croom} et~al.,}{{Croom} et~al.}{2012}]{SAMI}
{Croom} S.~M.,  et~al., 2012, \mn@doi [\mnras]
  {10.1111/j.1365-2966.2011.20365.x}, \href
  {http://adsabs.harvard.edu/abs/2012MNRAS.421..872C} {421, 872}

\bibitem[\protect\citeauthoryear{{Dale} et~al.,}{{Dale}
  et~al.}{2012}]{Dale2012}
{Dale} D.~A.,  et~al., 2012, \mn@doi [\apj] {10.1088/0004-637X/745/1/95}, \href
  {http://adsabs.harvard.edu/abs/2012ApJ...745...95D} {745, 95}

\bibitem[\protect\citeauthoryear{{Davis} et~al.,}{{Davis}
  et~al.}{2011}]{Davis2011}
{Davis} T.~A.,  et~al., 2011, \mn@doi [\mnras]
  {10.1111/j.1365-2966.2011.19355.x}, \href
  {http://adsabs.harvard.edu/abs/2011MNRAS.417..882D} {417, 882}

\bibitem[\protect\citeauthoryear{{De Looze}, {Baes}, {Fritz}  \&
  {Verstappen}}{{De Looze} et~al.}{2012}]{DeLooze2012a}
{De Looze} I.,  {Baes} M.,  {Fritz} J.,   {Verstappen} J.,  2012, \mn@doi
  [\mnras] {10.1111/j.1365-2966.2011.19759.x}, \href
  {http://adsabs.harvard.edu/abs/2012MNRAS.419..895D} {419, 895}

\bibitem[\protect\citeauthoryear{{Disney}, {Davies}  \& {Phillipps}}{{Disney}
  et~al.}{1989}]{Disney1989}
{Disney} M.,  {Davies} J.,   {Phillipps} S.,  1989, \mn@doi [\mnras]
  {10.1093/mnras/239.3.939}, \href
  {http://adsabs.harvard.edu/abs/1989MNRAS.239..939D} {239, 939}

\bibitem[\protect\citeauthoryear{{Dom{\'{\i}}nguez} et~al.,}{{Dom{\'{\i}}nguez}
  et~al.}{2013}]{Dominguez2013}
{Dom{\'{\i}}nguez} A.,  et~al., 2013, \mn@doi [\apj]
  {10.1088/0004-637X/763/2/145}, \href
  {http://adsabs.harvard.edu/abs/2013ApJ...763..145D} {763, 145}

\bibitem[\protect\citeauthoryear{{Dong} et~al.,}{{Dong}
  et~al.}{2014}]{Dong2014}
{Dong} H.,  et~al., 2014, \mn@doi [\apj] {10.1088/0004-637X/785/2/136}, \href
  {http://adsabs.harvard.edu/abs/2014ApJ...785..136D} {785, 136}

\bibitem[\protect\citeauthoryear{{Draine}}{{Draine}}{2003}]{Draine2003}
{Draine} B.~T.,  2003, \mn@doi [\araa]
  {10.1146/annurev.astro.41.011802.094840}, \href
  {http://adsabs.harvard.edu/abs/2003ARA%26A..41..241D} {41, 241}

\bibitem[\protect\citeauthoryear{{Eales}, {de Vis}, {W.~L.~Smith}, {Appah},
  {Ciesla}, {Duffield}  \& {Schofield}}{{Eales} et~al.}{2017}]{Eales2017}
{Eales} S.,  {de Vis} P.,  {W.~L.~Smith} M.,  {Appah} K.,  {Ciesla} L.,
  {Duffield} C.,   {Schofield} S.,  2017, \mn@doi [\mnras]
  {10.1093/mnras/stw2875}, \href
  {http://adsabs.harvard.edu/abs/2017MNRAS.465.3125E} {465, 3125}

\bibitem[\protect\citeauthoryear{{Ferrarese} et~al.,}{{Ferrarese}
  et~al.}{2006}]{Ferrarese2006}
{Ferrarese} L.,  et~al., 2006, \mn@doi [\apjs] {10.1086/501350}, \href
  {http://adsabs.harvard.edu/abs/2006ApJS..164..334F} {164, 334}

\bibitem[\protect\citeauthoryear{{Finkelman} et~al.,}{{Finkelman}
  et~al.}{2008}]{Finkelman2008}
{Finkelman} I.,  et~al., 2008, \mn@doi [\mnras]
  {10.1111/j.1365-2966.2008.13785.x}, \href
  {http://adsabs.harvard.edu/abs/2008MNRAS.390..969F} {390, 969}

\bibitem[\protect\citeauthoryear{{Finkelman} et~al.,}{{Finkelman}
  et~al.}{2010}]{Finkelman2010}
{Finkelman} I.,  et~al., 2010, \mn@doi [\mnras]
  {10.1111/j.1365-2966.2010.17334.x}, \href
  {http://adsabs.harvard.edu/abs/2010MNRAS.409..727F} {409, 727}

\bibitem[\protect\citeauthoryear{{Fitzpatrick}}{{Fitzpatrick}}{1999}]{Fitzpatrick1999}
{Fitzpatrick} E.~L.,  1999, \mn@doi [\pasp] {10.1086/316293}, \href
  {http://adsabs.harvard.edu/abs/1999PASP..111...63F} {111, 63}

\bibitem[\protect\citeauthoryear{{Fitzpatrick} \& {Massa}}{{Fitzpatrick} \&
  {Massa}}{2007}]{Fitzpatrick2007}
{Fitzpatrick} E.~L.,  {Massa} D.,  2007, \mn@doi [\apj] {10.1086/518158}, \href
  {http://adsabs.harvard.edu/abs/2007ApJ...663..320F} {663, 320}

\bibitem[\protect\citeauthoryear{{Gadotti}, {Baes}  \& {Falony}}{{Gadotti}
  et~al.}{2010}]{Gadotti2010}
{Gadotti} D.~A.,  {Baes} M.,   {Falony} S.,  2010, \mn@doi [\mnras]
  {10.1111/j.1365-2966.2010.16243.x}, \href
  {http://adsabs.harvard.edu/abs/2010MNRAS.403.2053G} {403, 2053}

\bibitem[\protect\citeauthoryear{{Gilbank}, {Baldry}, {Balogh}, {Glazebrook}
  \& {Bower}}{{Gilbank} et~al.}{2010}]{Gilbank2010}
{Gilbank} D.~G.,  {Baldry} I.~K.,  {Balogh} M.~L.,  {Glazebrook} K.,   {Bower}
  R.~G.,  2010, \mn@doi [\mnras] {10.1111/j.1365-2966.2010.16640.x}, \href
  {http://adsabs.harvard.edu/abs/2010MNRAS.405.2594G} {405, 2594}

\bibitem[\protect\citeauthoryear{{Gordon}, {Clayton}, {Misselt}, {Landolt}  \&
  {Wolff}}{{Gordon} et~al.}{2003}]{Gordon2003}
{Gordon} K.~D.,  {Clayton} G.~C.,  {Misselt} K.~A.,  {Landolt} A.~U.,   {Wolff}
  M.~J.,  2003, \mn@doi [\apj] {10.1086/376774}, \href
  {http://adsabs.harvard.edu/abs/2003ApJ...594..279G} {594, 279}

\bibitem[\protect\citeauthoryear{{Goudfrooij} \& {de Jong}}{{Goudfrooij} \& {de
  Jong}}{1995}]{Goudfrooij1995}
{Goudfrooij} P.,  {de Jong} T.,  1995, \aap, \href
  {http://adsabs.harvard.edu/abs/1995A%26A...298..784G} {298, 784}

\bibitem[\protect\citeauthoryear{{Goudfrooij}, {de Jong}, {Hansen}  \&
  {Norgaard-Nielsen}}{{Goudfrooij} et~al.}{1994}]{Goudfrooij1994b}
{Goudfrooij} P.,  {de Jong} T.,  {Hansen} L.,   {Norgaard-Nielsen} H.~U.,
  1994, \mnras, \href {http://adsabs.harvard.edu/abs/1994MNRAS.271..833G} {271,
  833}

\bibitem[\protect\citeauthoryear{{Groves}, {Brinchmann}  \& {Walcher}}{{Groves}
  et~al.}{2012}]{Groves2012}
{Groves} B.,  {Brinchmann} J.,   {Walcher} C.~J.,  2012, \mn@doi [\mnras]
  {10.1111/j.1365-2966.2011.19796.x}, \href
  {http://adsabs.harvard.edu/abs/2012MNRAS.419.1402G} {419, 1402}

\bibitem[\protect\citeauthoryear{{Hawarden}, {Longmore}, {Tritton}, {Elson}  \&
  {Corwin}}{{Hawarden} et~al.}{1981}]{Hawarden1981}
{Hawarden} T.~G.,  {Longmore} A.~J.,  {Tritton} S.~B.,  {Elson} R.~A.~W.,
  {Corwin} Jr. H.~G.,  1981, \mn@doi [\mnras] {10.1093/mnras/196.4.747}, \href
  {http://adsabs.harvard.edu/abs/1981MNRAS.196..747H} {196, 747}

\bibitem[\protect\citeauthoryear{{Herbig}}{{Herbig}}{1995}]{Herbig1995}
{Herbig} G.~H.,  1995, \mn@doi [\araa] {10.1146/annurev.aa.33.090195.000315},
  \href {http://adsabs.harvard.edu/abs/1995ARA%26A..33...19H} {33, 19}

\bibitem[\protect\citeauthoryear{{Jones}, {Fanciullo}, {K{\"o}hler},
  {Verstraete}, {Guillet}, {Bocchio}  \& {Ysard}}{{Jones}
  et~al.}{2013}]{Jones2013}
{Jones} A.~P.,  {Fanciullo} L.,  {K{\"o}hler} M.,  {Verstraete} L.,  {Guillet}
  V.,  {Bocchio} M.,   {Ysard} N.,  2013, \mn@doi [\aap]
  {10.1051/0004-6361/201321686}, \href
  {http://adsabs.harvard.edu/abs/2013A%26A...558A..62J} {558, A62}

\bibitem[\protect\citeauthoryear{{Jones}, {Koehler}, {Ysard}, {Bocchio}  \&
  {Verstraete}}{{Jones} et~al.}{2017}]{Jones2017}
{Jones} A.~P.,  {Koehler} M.,  {Ysard} N.,  {Bocchio} M.,   {Verstraete} L.,
  2017, preprint, \href {http://adsabs.harvard.edu/abs/2017arXiv170300775J} {}
  (\mn@eprint {arXiv} {1703.00775})

\bibitem[\protect\citeauthoryear{{Kreckel} et~al.,}{{Kreckel}
  et~al.}{2013}]{Kreckel2013}
{Kreckel} K.,  et~al., 2013, \mn@doi [\apj] {10.1088/0004-637X/771/1/62}, \href
  {http://adsabs.harvard.edu/abs/2013ApJ...771...62K} {771, 62}

\bibitem[\protect\citeauthoryear{{Kriek} \& {Conroy}}{{Kriek} \&
  {Conroy}}{2013}]{Kriek2013}
{Kriek} M.,  {Conroy} C.,  2013, \mn@doi [\apjl] {10.1088/2041-8205/775/1/L16},
  \href {http://adsabs.harvard.edu/abs/2013ApJ...775L..16K} {775, L16}

\bibitem[\protect\citeauthoryear{{Kr{\"u}gel}}{{Kr{\"u}gel}}{2009}]{Krugel2009}
{Kr{\"u}gel} E.,  2009, \mn@doi [\aap] {10.1051/0004-6361:200809976}, \href
  {http://adsabs.harvard.edu/abs/2009A%26A...493..385K} {493, 385}

\bibitem[\protect\citeauthoryear{{Kulkarni}, {Sahu}, {Chaware}, {Chakradhari}
  \& {Pandey}}{{Kulkarni} et~al.}{2014}]{Kulkarni2014}
{Kulkarni} S.,  {Sahu} D.~K.,  {Chaware} L.,  {Chakradhari} N.~K.,   {Pandey}
  S.~K.,  2014, \mn@doi [\na] {10.1016/j.newast.2014.01.003}, \href
  {http://adsabs.harvard.edu/abs/2014NewA...30...51K} {30, 51}

\bibitem[\protect\citeauthoryear{{Liu} et~al.,}{{Liu} et~al.}{2013}]{Liu2013}
{Liu} G.,  et~al., 2013, \mn@doi [\apjl] {10.1088/2041-8205/778/2/L41}, \href
  {http://adsabs.harvard.edu/abs/2013ApJ...778L..41L} {778, L41}

\bibitem[\protect\citeauthoryear{{Mathis}}{{Mathis}}{1990}]{Mathis1990}
{Mathis} J.~S.,  1990, \mn@doi [\araa] {10.1146/annurev.aa.28.090190.000345},
  \href {http://adsabs.harvard.edu/abs/1990ARA%26A..28...37M} {28, 37}

\bibitem[\protect\citeauthoryear{{Michard}}{{Michard}}{2005}]{Michard2005}
{Michard} R.,  2005, \mn@doi [\aap] {10.1051/0004-6361:20041955}, \href
  {http://adsabs.harvard.edu/abs/2005A%26A...441..451M} {441, 451}

\bibitem[\protect\citeauthoryear{{M{\"o}llenhoff}, {Popescu}  \&
  {Tuffs}}{{M{\"o}llenhoff} et~al.}{2006}]{Mollenhoff2006}
{M{\"o}llenhoff} C.,  {Popescu} C.~C.,   {Tuffs} R.~J.,  2006, \mn@doi [\aap]
  {10.1051/0004-6361:20054727}, \href
  {http://adsabs.harvard.edu/abs/2006A%26A...456..941M} {456, 941}

\bibitem[\protect\citeauthoryear{{Monreal-Ibero}, {Weilbacher}, {Wendt},
  {Selman}, {Lallement}, {Brinchmann}, {Kamann}  \& {Sandin}}{{Monreal-Ibero}
  et~al.}{2015}]{MonrealIbero2015}
{Monreal-Ibero} A.,  {Weilbacher} P.~M.,  {Wendt} M.,  {Selman} F.,
  {Lallement} R.,  {Brinchmann} J.,  {Kamann} S.,   {Sandin} C.,  2015, \mn@doi
  [\aap] {10.1051/0004-6361/201525854}, \href
  {http://adsabs.harvard.edu/abs/2015A%26A...576L...3M} {576, L3}

\bibitem[\protect\citeauthoryear{{Natta} \& {Panagia}}{{Natta} \&
  {Panagia}}{1984}]{Natta1984}
{Natta} A.,  {Panagia} N.,  1984, \mn@doi [\apj] {10.1086/162681}, \href
  {http://adsabs.harvard.edu/abs/1984ApJ...287..228N} {287, 228}

\bibitem[\protect\citeauthoryear{{Nelson} et~al.,}{{Nelson}
  et~al.}{2016}]{Nelson2016}
{Nelson} E.~J.,  et~al., 2016, \mn@doi [\apjl] {10.3847/2041-8205/817/1/L9},
  \href {http://adsabs.harvard.edu/abs/2016ApJ...817L...9N} {817, L9}

\bibitem[\protect\citeauthoryear{{Noll}, {Burgarella}, {Giovannoli}, {Buat},
  {Marcillac}  \& {Mu{\~n}oz-Mateos}}{{Noll} et~al.}{2009}]{Noll2009}
{Noll} S.,  {Burgarella} D.,  {Giovannoli} E.,  {Buat} V.,  {Marcillac} D.,
  {Mu{\~n}oz-Mateos} J.~C.,  2009, \mn@doi [\aap]
  {10.1051/0004-6361/200912497}, \href
  {http://adsabs.harvard.edu/abs/2009A%26A...507.1793N} {507, 1793}

\bibitem[\protect\citeauthoryear{{Osterbrock}}{{Osterbrock}}{1989}]{Osterbrock1989}
{Osterbrock} D.~E.,  1989, {Astrophysics of gaseous nebulae and active galactic
  nuclei}

\bibitem[\protect\citeauthoryear{{Patil}, {Pandey}, {Sahu}  \&
  {Kembhavi}}{{Patil} et~al.}{2007}]{Patil2007}
{Patil} M.~K.,  {Pandey} S.~K.,  {Sahu} D.~K.,   {Kembhavi} A.,  2007, \mn@doi
  [\aap] {10.1051/0004-6361:20053512}, \href
  {http://adsabs.harvard.edu/abs/2007A%26A...461..103P} {461, 103}

\bibitem[\protect\citeauthoryear{{Popescu}, {Tuffs}, {V{\"o}lk}, {Pierini}  \&
  {Madore}}{{Popescu} et~al.}{2002}]{Popescu2002}
{Popescu} C.~C.,  {Tuffs} R.~J.,  {V{\"o}lk} H.~J.,  {Pierini} D.,   {Madore}
  B.~F.,  2002, \mn@doi [\apj] {10.1086/338383}, \href
  {http://adsabs.harvard.edu/abs/2002ApJ...567..221P} {567, 221}

\bibitem[\protect\citeauthoryear{{Poznanski}, {Ganeshalingam}, {Silverman}  \&
  {Filippenko}}{{Poznanski} et~al.}{2011}]{Poznanski2011}
{Poznanski} D.,  {Ganeshalingam} M.,  {Silverman} J.~M.,   {Filippenko} A.~V.,
  2011, \mn@doi [\mnras] {10.1111/j.1745-3933.2011.01084.x}, \href
  {http://adsabs.harvard.edu/abs/2011MNRAS.415L..81P} {415, L81}

\bibitem[\protect\citeauthoryear{{Poznanski}, {Prochaska}  \&
  {Bloom}}{{Poznanski} et~al.}{2012}]{Poznanski2012}
{Poznanski} D.,  {Prochaska} J.~X.,   {Bloom} J.~S.,  2012, \mn@doi [\mnras]
  {10.1111/j.1365-2966.2012.21796.x}, \href
  {http://adsabs.harvard.edu/abs/2012MNRAS.426.1465P} {426, 1465}

\bibitem[\protect\citeauthoryear{{Reddy} et~al.,}{{Reddy}
  et~al.}{2015}]{Reddy2015}
{Reddy} N.~A.,  et~al., 2015, \mn@doi [\apj] {10.1088/0004-637X/806/2/259},
  \href {http://adsabs.harvard.edu/abs/2015ApJ...806..259R} {806, 259}

\bibitem[\protect\citeauthoryear{{Sahu}, {Pandey}  \& {Kembhavi}}{{Sahu}
  et~al.}{1998}]{Sahu1998}
{Sahu} D.~K.,  {Pandey} S.~K.,   {Kembhavi} A.,  1998, \aap, \href
  {http://adsabs.harvard.edu/abs/1998A%26A...333..803S} {333, 803}

\bibitem[\protect\citeauthoryear{{Salmon} et~al.,}{{Salmon}
  et~al.}{2016}]{Salmon2016}
{Salmon} B.,  et~al., 2016, \mn@doi [\apj] {10.3847/0004-637X/827/1/20}, \href
  {http://adsabs.harvard.edu/abs/2016ApJ...827...20S} {827, 20}

\bibitem[\protect\citeauthoryear{{Sarre}}{{Sarre}}{2006}]{Sarre2006}
{Sarre} P.~J.,  2006, \mn@doi [Journal of Molecular Spectroscopy]
  {10.1016/j.jms.2006.03.009}, \href
  {http://adsabs.harvard.edu/abs/2006JMoSp.238....1S} {238, 1}

\bibitem[\protect\citeauthoryear{{Sarzi} et~al.,}{{Sarzi}
  et~al.}{2006}]{Sarzi2006}
{Sarzi} M.,  et~al., 2006, \mn@doi [\mnras] {10.1111/j.1365-2966.2005.09839.x},
  \href {http://adsabs.harvard.edu/abs/2006MNRAS.366.1151S} {366, 1151}

\bibitem[\protect\citeauthoryear{{Scicluna} \& {Siebenmorgen}}{{Scicluna} \&
  {Siebenmorgen}}{2015}]{Scicluna2015}
{Scicluna} P.,  {Siebenmorgen} R.,  2015, \mn@doi [\aap]
  {10.1051/0004-6361/201323149}, \href
  {http://adsabs.harvard.edu/abs/2015A%26A...584A.108S} {584, A108}

\bibitem[\protect\citeauthoryear{{Seon} \& {Draine}}{{Seon} \&
  {Draine}}{2016}]{Seon2016}
{Seon} K.-I.,  {Draine} B.~T.,  2016, \mn@doi [\apj]
  {10.3847/1538-4357/833/2/201}, \href
  {http://adsabs.harvard.edu/abs/2016ApJ...833..201S} {833, 201}

\bibitem[\protect\citeauthoryear{{Skibba} et~al.,}{{Skibba}
  et~al.}{2011}]{Skibba2011}
{Skibba} R.~A.,  et~al., 2011, \mn@doi [\apj] {10.1088/0004-637X/738/1/89},
  \href {http://adsabs.harvard.edu/abs/2011ApJ...738...89S} {738, 89}

\bibitem[\protect\citeauthoryear{{Smith} et~al.,}{{Smith}
  et~al.}{2012}]{Smith2012b}
{Smith} M.~W.~L.,  et~al., 2012, \mn@doi [\apj] {10.1088/0004-637X/756/1/40},
  \href {http://adsabs.harvard.edu/abs/2012ApJ...756...40S} {756, 40}

\bibitem[\protect\citeauthoryear{{Snow}}{{Snow}}{2014}]{Snow2014}
{Snow} T.~P.,  2014, in {Cami} J.,  {Cox} N.~L.~J.,  eds,  IAU Symposium Vol.
  297, The Diffuse Interstellar Bands. pp 3--12,
  \mn@doi{10.1017/S1743921313015512}

\bibitem[\protect\citeauthoryear{{Valencic}, {Clayton}  \& {Gordon}}{{Valencic}
  et~al.}{2004}]{Valencic2004}
{Valencic} L.~A.,  {Clayton} G.~C.,   {Gordon} K.~D.,  2004, \mn@doi [\apj]
  {10.1086/424922}, \href {http://adsabs.harvard.edu/abs/2004ApJ...616..912V}
  {616, 912}

\bibitem[\protect\citeauthoryear{{Vanderbeke}, {Baes}, {Romanowsky}  \&
  {Schmidtobreick}}{{Vanderbeke} et~al.}{2011}]{Vanderbeke2011}
{Vanderbeke} J.,  {Baes} M.,  {Romanowsky} A.~J.,   {Schmidtobreick} L.,  2011,
  \mn@doi [\mnras] {10.1111/j.1365-2966.2010.18036.x}, \href
  {http://adsabs.harvard.edu/abs/2011MNRAS.412.2017V} {412, 2017}

\bibitem[\protect\citeauthoryear{{Viaene} et~al.,}{{Viaene}
  et~al.}{2015}]{Viaene2015}
{Viaene} S.,  et~al., 2015, \mn@doi [\aap] {10.1051/0004-6361/201526147}, \href
  {http://adsabs.harvard.edu/abs/2015A%26A...579A.103V} {579, A103}

\bibitem[\protect\citeauthoryear{{Viaene} et~al.,}{{Viaene}
  et~al.}{2016}]{Viaene2016}
{Viaene} S.,  et~al., 2016, \mn@doi [\aap] {10.1051/0004-6361/201527586}, \href
  {http://adsabs.harvard.edu/abs/2016A%26A...586A..13V} {586, A13}

\bibitem[\protect\citeauthoryear{{Wild}, {Charlot}, {Brinchmann}, {Heckman},
  {Vince}, {Pacifici}  \& {Chevallard}}{{Wild} et~al.}{2011}]{Wild2011}
{Wild} V.,  {Charlot} S.,  {Brinchmann} J.,  {Heckman} T.,  {Vince} O.,
  {Pacifici} C.,   {Chevallard} J.,  2011, \mn@doi [\mnras]
  {10.1111/j.1365-2966.2011.19367.x}, \href
  {http://adsabs.harvard.edu/abs/2011MNRAS.417.1760W} {417, 1760}

\bibitem[\protect\citeauthoryear{{Ysard}, {K{\"o}hler}, {Jones},
  {Miville-Desch{\^e}nes}, {Abergel}  \& {Fanciullo}}{{Ysard}
  et~al.}{2015}]{Ysard2015}
{Ysard} N.,  {K{\"o}hler} M.,  {Jones} A.,  {Miville-Desch{\^e}nes} M.-A.,
  {Abergel} A.,   {Fanciullo} L.,  2015, \mn@doi [\aap]
  {10.1051/0004-6361/201425523}, \href
  {http://adsabs.harvard.edu/abs/2015A%26A...577A.110Y} {577, A110}

\makeatother
\end{thebibliography}




\bsp	
\label{lastpage}
\end{document}